\documentclass{aa}  

\usepackage{epstopdf}
\usepackage{mathtools}
\usepackage{mathrsfs}
\usepackage{calrsfs}
\usepackage{natbib}

\usepackage[autostyle, english = american]{csquotes}
\MakeOuterQuote{"}
\usepackage{bm}
\usepackage{graphicx}
\usepackage{txfonts}
\usepackage{placeins}
\usepackage{url}
\usepackage{hyperref}
\usepackage{enumitem}

\hypersetup{
    colorlinks,
    linkcolor={blue},
    citecolor={blue},
    urlcolor={blue}
}

 \usepackage{lineno}

\begin{document}

\newcommand{\oiii}{\mbox{O\,{\sc iii}}}

\title{Jet Collimation and Acceleration in the Flat Spectrum Radio Quasar 1928+738}

    \author{Kunwoo Yi\inst{1}, 
           Jongho Park\inst{2}, 
           Masanori Nakamura\inst{3,4},
           Kazuhiro Hada\inst{5,6},
           \and
           Sascha Trippe\inst{1,7}
           }

    \institute{Department of Physics and Astronomy, Seoul National University, Gwanak-gu, Seoul 08826, Republic of Korea\\
              \email{kunwoo.yi@snu.ac.kr, jparkastro@khu.ac.kr}
        \and
            Department of Astronomy and Space Science, Kyung Hee University, 1732, Deogyeong-daero, Giheung-gu, Yongin-si, Gyeonggi-do 17104, Republic of Korea
        \and
            National Institute of Technology, Hachinohe College, 16-1 Uwanotai, Tamonoki, Hachinohe, Aomori 039-1192, Japan            
        \and
            Institute of Astronomy and Astrophysics, Academia Sinica, P.O. Box 23-141, Taipei 10617, Taiwan
        \and
            Mizusawa VLBI Observatory, National Astronomical Observatory of Japan, Osawa, Mitaka, Tokyo 181-8588, Japan
        \and
            Department of Astronomical Science, The Graduate University for Advanced Studies (SOKENDAI), 2-21-1 Osawa, Mitaka, Tokyo 181-8588, Japan
        \and
            SNU Astronomy Research Center, Seoul National University, Gwanak-gu, Seoul 08826, Republic of Korea
             }
   \date{Received \today; accepted \today}

  \abstract
  {Using time-resolved multifrequency Very Long Baseline Array data and new KaVA (KVN and VERA Array) observations, we study the structure and kinematics of the jet of the flat spectrum radio quasar (FSRQ) 1928+738. We find two distinct jet geometries as function of distance from the central black hole, with the inner jet having a parabolic shape, indicating collimation, and the outer jet having a conical shape, indicating free expansion of the jet plasma. Jet component speeds display a gradual outward acceleration up to a bulk Lorentz factor $\Gamma_{\rm max} \approx10$, followed by a deceleration further downstream. The location of the acceleration zone matches the region where the jet collimation occurs; this is the first direct observation of an acceleration and collimation zone (ACZ) in an FSRQ. The ACZ terminates approximately at a distance of 5.6\,$\times\,10^6$ gravitational radii, which is in good agreement with the sphere of gravitational influence of the supermassive black hole, implying that the physical extent of the ACZ is controlled by the black hole gravity. Our results suggest that confinement by an external medium is responsible for the jet collimation and that the jet is accelerated by converting Poynting flux energy to kinetic energy. 
  }

   \titlerunning{Jet Collimation and Acceleration in the Flat Spectrum Radio Quasar 1928+738}
   \authorrunning{Yi et al.}
   
   \maketitle
%

\section{Introduction}
\label{sec:intro}

A fraction ($\sim10\%$) of active galactic nuclei (AGNs) has highly collimated relativistic jets \citep{Urry1995}. It is now widely believed that these jets are launched in the vicinity of supermassive black holes (SMBHs) at the center of their host galaxies. An interplay of magnetic field, accreted gas, and rotation of the black hole itself \citep{Blandford1977} and/or the accretion disk \citep{Blandford1982} will produce the momentum and energy required for the formation of jets. The parsec and kilo-parsec scale properties of AGN jets have been intensively studied theoretically and observationally during more than five decades \citep[see][for a review]{Blandford2019, Hada2019}. 

A magnetically driven jet \citep{Blandford1977, Blandford1982} is expected to be accelerated to highly relativistic speeds by a magnetohydrodynamic (MHD) conversion of Poynting flux to kinetic energy \citep[e.g.,][]{Vlahakis2004}. This process is efficient if the bulk flow and the poloidal magnetic field lines therein are collimated by the magnetic nozzle effect \citep[e.g.,][]{Camenzind1987, Li1992, Begelman1994, Vlahakis2015}, implying that the bulk jet acceleration is intimately associated with jet collimation. It has been suggested that a combined acceleration and collimation zone (ACZ) is formed at distances $\lesssim 10^5$--$10^6$ gravitational radii ($R_{g}$) from the BH \citep[e.g.,][]{Meier2001, Komissarov2007, Marscher2008}. Numerous high-resolution observations with Very Long Baseline Interferometry (VLBI) have been performed toward AGN jets to examine the ACZ, which requires resolving the jet structure over a wide range of physical scales. 

The existence of an ACZ was confirmed in the nearby radio galaxy M\,87; \citet{Junor1999} found that the M\,87 jet is being collimated at a distance of $\lesssim$ 10 pc, and \citet{Asada2012} reported that the jet collimation is sustained up to $\sim 10^5\,R_{g}$, breaking at around the Bondi radius. This suggests that the gas accreted onto the central black hole, which is stratified by the black hole's gravity, might be responsible for shaping a relativistic AGN jet into a parabolic stream. This interpretation is consistent with the theoretical expectation that a jet requires confinement by an ambient medium to be collimated \citep[e.g.,][]{Lyubarsky2009}. Likewise, the acceleration of the jet takes place inside the Bondi radius \citep{Nakamura2013, Asada2014, Mertens2016, Hada2017, Walker2018, Park2019-Kine}, transiting into a slow deceleration beyond the Bondi radius \citep{Biretta1995,Biretta1999, Meyer2013}.

Our understanding of jet collimation has evolved greatly with the jets of other radio-loud AGNs being extensively investigated in the same manner. An active collimation at $\lesssim 10^4\,R_{g}$ \citep[e.g.,][]{Giovannini2018, Boccardi2019}, and a transition into a free expansion phase further downstream \citep[see][]{Tseng2016, Akiyama2018, Nakahara2018, Hada2018, Nakahara2020, Traianou2020, Kovalev2020, Park2021, Boccardi2021, Burd2022, Okino2022} have been discovered in an increasing number of AGN jets. 

Even though, jet acceleration within the collimation zone is mostly unexplored except for a few sources. Hence, several key questions such as `how and where is the bulk acceleration terminated' or `what is the maximum speed $\rm \Gamma_{max}$' remain unclear. This is partly due to the lack of monitoring observations. Indeed, a robust analysis of the jet velocity field requires multi-epoch VLBI observations at a high cadence. To date, jet acceleration and collimation in the same region, and thus a genuine ACZ, has been observed only in two Fanaroff-Riley type I \citep[FR I;][]{Fanaroff1974} radio galaxies (\citealp[M\,87;][]{Asada2012, Asada2014} and \citealp[NGC\,315;][]{Park2021, Boccardi2021, Ricci2022}) and one narrow-line Seyfert\,1 (NLS1) galaxy (1H\,0323+342; \citealt{Hada2018}). Notably, while the sample size has increased slowly, there has been no detailed study of the interplay of jet acceleration and collimation in a flat spectrum radio quasar (FSRQ), the FR II-like AGN sub-class, so far.

FSRQ 1928+738 (a.k.a. 4C\,+73.18) is located at a redshift of 0.302 \citep{Lawrence1986} and hosts a SMBH with a mass of $M_{\bullet}\,\approx\,3.7 \times 10^{8}$ $M_{\odot}$ in its center \citep[see][and the references therein]{Park2017}. It is noteworthy that 1928+738 is one of very few `misaligned' FSRQs with an unusually large viewing angle of $\theta_{\rm v}\,\simeq\,10^{\circ}$--$15^{\circ}$ between the jet and the line of sight \citep{Lahteenmaki1999, Hovatta2009, Liodakis2018}. Adopting $\theta_{\rm v}\,=\,13.2^{\circ}$ (see Section\,\ref{subsec:KaVA_var}), one finds a convenient conversion between angular scale and deprojected physical scale: 1\,mas $\sim\,10^6\,R_{g}$. Therefore, 1928+738 is one of the best FSRQ targets to directly investigate its putative jet ACZ with (sub)-mas VLBI observations. 

While Very Large Array (VLA)\,/\,MERLIN observations show a two-sided jet and structure on arcsecond (and thus kpc) scales \citep{Johnston1987, Hummel1992}, Very Long Baseline Array (VLBA) images show only a one-sided jet extending to the south \citep{Eckart1985}. At parsec scales, it has long been known that the 1928+738 jet shows apparently superluminal motion. Interestingly, previous well-sampled observations \citep[e.g.,][]{Homan2001, Kellermann2004} found a hint of a gradual increase in jet speed versus distance. Several studies \citep[e.g.,][]{Kun2014, Homan2015, Lister2021} find consistent maximum proper motions ($\beta_{\rm app} \approx 8$), but disagree on the location of the fastest jet speed. The suggested location spans from $\approx 1$ to $\approx 13$ mas from the origin of the jet. 

In this paper, we investigate the jet collimation and acceleration of 1928+738 in detail. We present the results of multifrequency VLBI observations. Based on single-epoch deep imaging and multi-epoch dense monitoring, we explored the structural evolution of the radio jet, its kinematics, and variability. A dedicated spectral analysis will be presented in a forthcoming paper. Our paper is structured as follows. In Section~\ref{sec:obs}, we describe archival data, our new observations, and the data reduction. In Sections~\ref{sec:analysis} and \ref{sec:discussion}, our results and analysis are presented and discussed. Finally, we summarize the results and implications of our study in Section~\ref{sec:summary}. Throughout this paper, we adopt a cosmology with the following parameters: $\rm H_{0} = 70 ~km~s^{-1}~Mpc^{-1}$, $\Omega_{m} = 0.3$, $\Omega_{\Lambda} = 0.7$. This results in a luminosity distance of 1565\,Mpc, and an angular-to-linear scale conversion of 4.47\,pc $\rm mas^{-1}$ in projection. Then, a proper motion of 1\,mas\,$\rm yr^{-1}$ translates to an apparent superluminal speed of 19\,c.

\section{Data and Data reduction}
\label{sec:obs}

In this section, we describe VLBI data taken from various public archives as well as new observations of FSRQ 1928+738. 

\subsection{VLBA archival data for jet collimation analysis}

We searched the National Radio Astronomy Observatory (NRAO) archive for VLBA data suitable for a study of jet collimation. We selected the VLBA project BS266 in which 1928+738 is observed over 10 hours, utilizing 9 VLBA stations except Saint Croix (VLBA-SC). The observations were made at 5 frequencies of 1.6\,GHz, 4\,GHz, 7\,GHz, 15\,GHz and 24\,GHz. We summarize the data in Table~\ref{table:VLBA}.

\begin{table}[t!]
\centering 
\caption{VLBA Project BS266 in 2018 June \label{table:VLBA}}
\begin{tabular}{c c c c}  
\hline
\hline 
Frequency & Synthesized Beam & $I_{\rm{peak}}$ & $I_{\rm{rms}}$ \\
(GHz) & (mas $\times$ mas, degree) & (Jy/bm) & (mJy/bm) \\
\hline
23.9 & 0.54 $\times$ 0.41, $-$9  & 2.44 & 0.70 \\
15.2 & 0.83 $\times$ 0.67, $-$9  & 3.25 & 0.53 \\
7.65 & 1.70 $\times$ 1.33, $-$14 & 3.46 & 0.78 \\
4.15 & 3.06 $\times$ 2.40, $-$15 & 2.28 & 0.76 \\
1.55 & 8.20 $\times$ 6.29, $-$12 & 1.71 & 0.46 \\ \hline
\end{tabular}
\tablefoot{Columns from \emph{left to right}: (1) observing frequency; (2) synthesized beam from naturally weighting; (3) peak intensity of the image; (4) rms noise of the image.}
\end{table}

\begin{table}[t!]
\centering 
\caption{Summary of Data Sets Used for Jet Kinematics \label{table:monitoring}}
\begin{tabular}{c c c c c}  
\hline
\hline 
Database & Band & Frequency & Typical Beam & Epochs  \\
\hline
\rm{Astrogeo} & \rm{S} & \rm{2.3\,GHz}        & 5.43 mas & 8  \\
\rm{Astrogeo} & \rm{X} & \rm{7.6 -- 8.6\,GHz} & 1.57 mas & 16 \\
\rm{MOJAVE}   & \rm{U} & \rm{15\,GHz}         & 0.63 mas & 14 \\
\rm{KaVA}     & \rm{Q} & \rm{43\,GHz}         & 0.69 mas & 22 \\ \hline
\end{tabular}
\tablefoot{Columns from \emph{left to right}: (1) name of the database; (2), (3) observing band and its central frequency; (4) average of the synthesized beam size; (5) number of observation epochs used in our analysis.}
\end{table}

\begin{table*}[t!]
\centering 
\caption{Summary of KaVA monitoring for 1928+738 at 43 GHz. \label{table:KaVA}}
\begin{tabular}{c c c c c c c c c c }  
\hline
\hline 
Code & Obs. date & $\rm t_{obs}$ & $\rm I_{peak}$ & $\sigma_{\rm rms}$ & $S_{\rm tot}$ & Synthesized Beam\\
  &  & (hour) & (Jy/beam) & (mJy/beam) & (Jy)  & (mas $\times$ mas,  degree) \\
\hline

r17045b          & 2017-02-14 & 8 & 1.74 & 0.73 & 2.51 & 0.77 $\times$ 0.73, $-$69.3 \\
r17073a          & 2017-03-14 & 8 & 2.09 & 1.17 & 3.07 & 0.74 $\times$ 0.66, $-$16.8\\
r17098a          & 2017-04-08 & 8 & 2.32 & 1.24 & 3.38 & 0.68 $\times$ 0.61, $-$20.6\\
r17129b          & 2017-05-09 & 8 & 2.40 & 1.45 & 3.49 & 0.67 $\times$ 0.56, $-$34.2\\
r17165a          & 2017-06-14 & 8 & 2.76 & 0.88 & 3.74 & 0.77 $\times$ 0.69, $-$34.8\\
r17249a          & 2017-09-06 & 6 & 2.50 & 1.51 & 3.47 & 0.70 $\times$ 0.52, $-$15.9\\
r17275a          & 2017-10-02 & 6 & 2.65 & 1.35 & 3.50 & 0.75 $\times$ 0.59, $-$19.2\\
r17290a          & 2017-10-17 & 6 & 2.46 & 1.02 & 3.43 & 0.71 $\times$ 0.57, 13.5\\
r17315a          & 2017-11-11 & 6 & 2.56 & 0.78 & 3.33 & 0.82 $\times$ 0.65, $-$14.8\\
r17344a          & 2017-12-10 & 6 & 2.14 & 1.96 & 2.80 & 0.84 $\times$ 0.57, $-$46.9\\
r18005a          & 2018-01-05 & 6 & 2.32 & 1.10 & 3.03 & 0.81 $\times$ 0.63, $-$19.8\\
r18054b          & 2018-02-23 & 8 & 2.26 & 0.88 & 3.05 & 0.76 $\times$ 0.64, $-$10.6\\
r18069b          & 2018-03-10 & 8 & 2.21 & 0.80 & 3.02 & 0.74 $\times$ 0.65, 5.8\\
r18101b          & 2018-04-11 & 8 & 2.41 & 1.32 & 3.27 & 0.74 $\times$ 0.66, $-$22.5\\
r18129a          & 2018-05-09 & 8 & 2.51 & 1.25 & 3.33 & 0.79 $\times$ 0.67, $-$14.6\\
r18154b          & 2018-06-03 & 8 & 2.57 & 1.18 & 3.35 & 0.78 $\times$ 0.66, $-$22.0\\
r18248a$^{*}$    & 2018-09-05 & 8 & 2.15 & 1.38 & 2.92 & 0.80 $\times$ 0.56, 1.41\\
r18283a          & 2018-10-10 & 8 & 1.89 & 0.97 & 2.62 & 0.71 $\times$ 0.65, $-$15.8\\
r18316a          & 2018-11-12 & 8 & 1.73 & 0.82 & 2.37 & 0.77 $\times$ 0.62, $-$14.6\\
r18344a          & 2018-12-10 & 8 & 1.66 & 0.70 & 2.36 & 0.77 $\times$ 0.67, $-$24.0\\
r18361b$^{\dag}$ & 2018-12-27 & 8 & 1.60 & 1.06 & 2.14 & 0.88 $\times$ 0.64, 5.78\\
r19014a          & 2019-01-14 & 8 & 1.36 & 0.66 & 2.13 & 0.83 $\times$ 0.65, $-$58.0\\ \hline
\multicolumn{7}{l}{$^{*}$ KaVA without KVN-Yonsei.}\\
\multicolumn{7}{l}{$^{\dag}$ KaVA without KVN-Ulsan and VERA-Mizusawa.}\\
\end{tabular}
\tablefoot{Columns from \emph{left to right}: (1) Project codes; (2) date of observation; (3) total observation time; (4) peak Intensity; (5) rms noise; (6) total flux density of the image; (7) synthesized beam from naturally weighting.}

\end{table*}

\subsection{VLBA archival data for jet kinematics analysis}

We obtained multi-epoch VLBI data from various archival databases, summarized in Table~\ref{table:monitoring}. The S- and X-band (2.3 and 7.6 -- 8.6\,GHz) data sets are from the Astrogeo VLBI FITS image database\footnote{\url{http://astrogeo.org/vlbi_images/}}, which contains extensive surveys such as several versions of the \emph{VLBA Calibrator Survey} \citep[VCS;][]{Petrov2006,Kovalev2007,Petrov2008, Petrov2021}, and the \emph{VLBI 2MASS Survey} \citep[V2M;][]{Condon2011}. In total, we analyzed the calibrated data observed over 8 and 16 epochs from 2005 to 2019 for S- and X- band, respectively. We refer the readers to \citet[][]{Pushkarev2012_astrogeo} for detailed descriptions of the data and the data processing.

The 15\,GHz data set is from the MOJAVE database\footnote{\url{https://www.cv.nrao.edu/MOJAVE/sourcepages/1928+738.shtml}} \citep[e.g.,][]{Lister2018}. We analyzed 14 epochs of calibrated visibility data spanning from 2016 January to 2019 August. We note that we only consider the total intensity, not the polarized emission.

\subsection{KaVA 43 GHz data}
In addition to the VLBA archival data mentioned above, we also performed a new dedicated observing program on 1928+738 at 43\,GHz. We conducted high cadence (approximately monthly) monitoring with the KVN\footnote{Korean VLBI Network, which consists of three 21\,m telescopes in Korea} and VERA\footnote{VLBI Exploration of Radio Astrometry, which consists of four 20\,m telescopes in Japan} Array \citep[KaVA;][]{Niinuma2014, Hada2017, Park2019-Kine, Cho2022}, which plays as a core array of the East Asian VLBI Network \citep[EAVN;][]{Wajima2016,An2018,Cui2023}. In each session, all seven stations were scheduled by default in a 6 -- 8 hour track. One or two stations were missing occasionally, due to local issues such as bad weather or system failures (see Table~\ref{table:KaVA}). On 2018 January 22, due to the loss of VERA-Mizusawa and KVN-Tamna stations, the data quality was severely compromised, leading to the exclusion of the data. In total, we obtained 22 epochs of data spanning from 2017 February to 2019 January. We summarize our KaVA observations in Table~\ref{table:KaVA}.

All observations used the single polarization (left-hand circular only) mode with 2-bit quantization. The data were recorded at a rate of 1\,Gbps in 8 sub-bands of 32\,MHz width each, and correlated at the Korea–Japan Correlation Center (KJCC) using the Daejeon Hardware Correlator \citep{Lee2015_kjcc}.

\subsection{Data Reduction \label{subsec:reduction}}

After correlation, the raw data were calibrated by using the NRAO Astronomical Image Processing System \citep[AIPS;][]{Greisen2003}. We applied the standard KaVA\,/\,VLBA data reduction procedures \citep{Niinuma2014, Oh2015, Hada2017, Kino2018, Lee2019, Park2019-Kine, Hada2020, Wajima2020} for the initial phase, bandpass and amplitude calibrations. A priori amplitude calibration was applied by using the gain curve and system temperature of each antenna. Notably, the VLBA data underwent a manual atmospheric opacity correction using the APCAL procedure in AIPS. This step was necessary only for VLBA data, as the KaVA system temperatures were already opacity-corrected during the observations. Additionally, to correct the multiple losses stemming from signal processing in the observing system and the characteristics in Daejeon correlator, the visibility amplitudes of the KaVA data were up-scaled by a factor of 1.3 \citep{Lee2015_amp135}. Iterative CLEAN and self-calibration process were performed in the Difmap software \citep{Shepherd1997} for imaging.

\section{Analysis and results}
\label{sec:analysis}

\begin{figure*}[ht!]
\centering
\includegraphics[width=\textwidth]{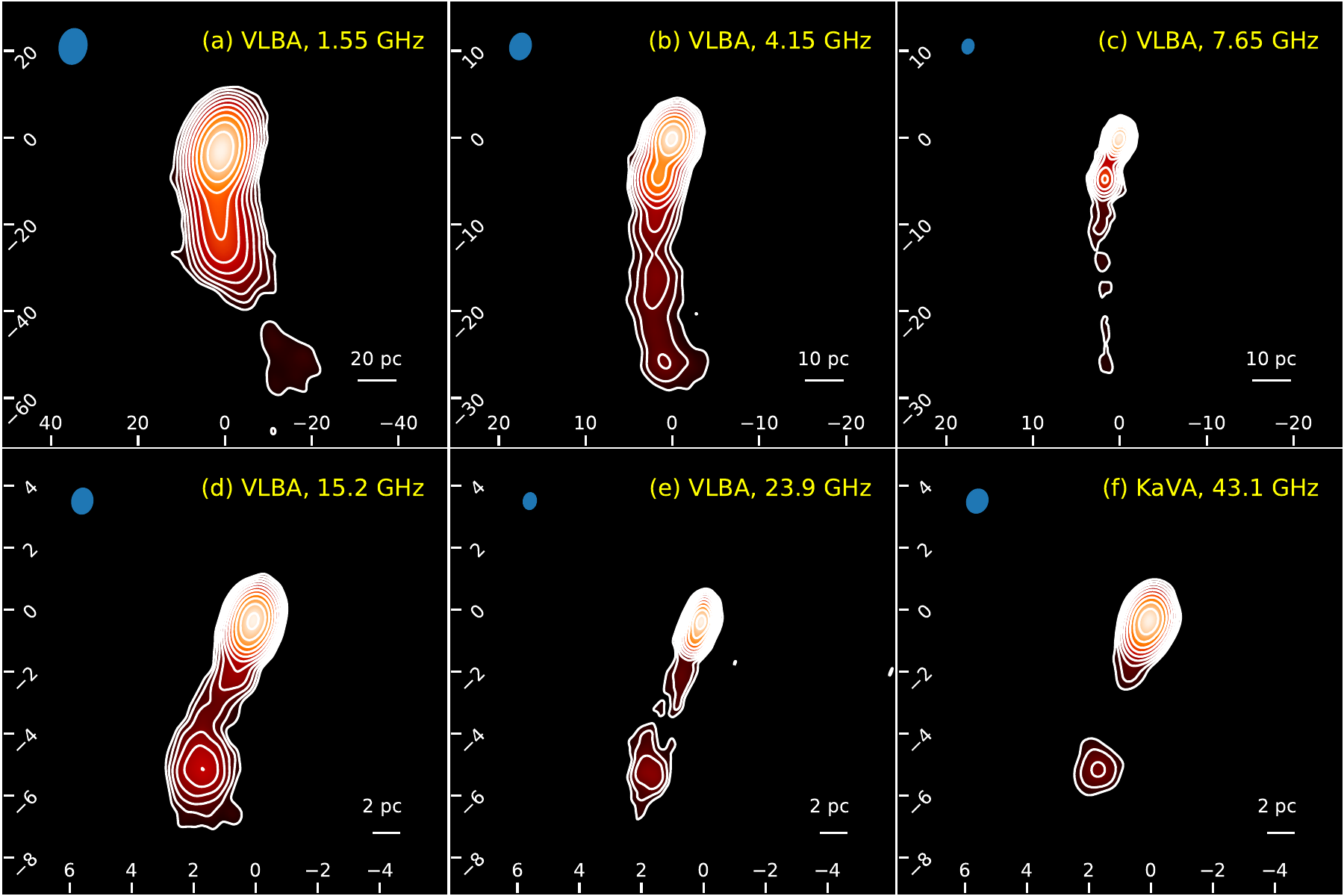}
    \caption{Naturally weighted CLEAN images of FSRQ 1928+738. at 1.55, 4.15, 7.65, 15.2, 23.9, and 43.1 GHz, respectively. For each image, contours start at the 5\,$\sigma$ image rms level and increase by factors of $2^{1/2}$. Physical scales are indicated in each map. Axis scales are in milli-arcsec. The blue ellipse in the top left corner of each map is the synthesized beam.}
    \label{Fig:imgs}
\end{figure*}

\subsection{Multifrequency Jet Images on Various Scales \label{subsec:images}}

We analyze the radial profiles of the width and velocity field of the jet of 1928+738 to investigate its collimation and acceleration. All displacements and directions are measured with respect to the ``core'' for analysis. Throughout this paper, we use the conventional label ``core'' for the innermost and compact feature of the radio jet. The core is presumably stationary and thus used as a reference point. We note that all images are shifted to place the core at the origin (RA, Dec)\,=\,(0, 0) of each map. Note however that the physical core position may depend on observing frequency (see Section~\ref{subsec:coreshift}).

We present VLBA and KaVA images of FSRQ 1928+738 obtained at six different frequencies in Figure~\ref{Fig:imgs}. The standard core--jet morphology is clearly observed at all frequencies, as are the individual jet structures on various scales.

At a frequency of 1.6\,GHz, the image displays an extended jet, covering radial distances down to approximately 40\,mas, followed by an isolated feature at around 60\,mas. This is consistent with previous VLBA observations at a similar frequency (1.4\,GHz) by \citet{Pushkarev2017}. The jet structure gradually becomes compact as the observing frequency increases, whereas the overall morphology remains broadly the same. Beyond the core region, the jet emission is dominated by an isolated feature at $\approx\,5$\,mas. This feature remains distinctly visible even at 43\,GHz, owing to its flat spectrum. We note that this feature is distinguishable from the core at all frequencies and can be identified via \emph{modelfit} even at the lowest resolution at 1.6\,GHz (see Section~\ref{subsec:coreshift}).

It has been previously reported that the 1928+738 jet exhibits curvature \citep[see][]{Eckart1985, Hummel1992, Roos1993, Homan2001, Homan2009, Kun2014, Roland2015}. Figure~\ref{Fig:imgs} shows a consistent result. The jet extends to the southeast with an approximate position angle (P.A.) of $160^{\circ}$, down to the bright feature at $\approx\,5$\,mas. Then, the jet direction gradually turns toward the southwest: the furthest structure at 1.6\,GHz has a P.A.~$\sim 195^{\circ}$. This behavior has been interpreted as a consequence of jet precession \citep{Roos1993, Kun2014, Roland2015} or a helical jet structure \citep{Hummel1992, Cheng2018}. \citet{Homan2001, Homan2009} suggested that the variation of position angle may result from jet collimation. We note that the origin of the jet curvature is beyond the scope of this work. Nevertheless, we have to consider the possibility that the viewing angle may not be constant even at the smallest scales \citep[see][]{Raiteri2017, Readhead2021}. We will discuss this issue in Section~\ref{subsec:KaVA_var}.

\subsection{Core Identification, Core Shift, Image Alignment\label{subsec:coreshift}}

The emission from the inner regions of AGN jets is usually optically thick because of synchrotron self-absorption and free-free absorption. Thus, the radio core is usually assumed to be a photosphere, i.e., a region where the optical depth of the jet plasma reaches unity \citep{Blandford1979, Konigl1981}. We used a standard model-fitting routine, the Difmap task \emph{modelfit}. After calibration and imaging, the visibility data at each frequency and each epoch was fitted with a set of circular Gaussian components. At any frequency, we identified the most upstream component as the core. The minimum resolvable size of Gaussian components is approximately $\rm 10\,\mu as$, assuming signal-to-noise ratios of several thousand \citep{Lobanov2005}. We note that the cores of 1928+738 appear to be unresolved even at the highest resolution, yielding a compactness of $\rm \theta_{core} \lesssim 10\,\mu as$ at any given frequency.

The absorption mechanism depends on the observing frequency \citep[e.g.,][]{Rybicki1979, Levinson1995}; so does the location of the core \citep[e.g.,][]{Marcaide1984, Lobanov1998, Hirotani2005}: The higher the observing frequency, the closer is the core to the central black hole -- the ``core-shift''. The opacity effect thereby prevents the usage of the core as a common reference point when combining multifrequency data. Alignment of the jet images at different frequencies should be preceded by a precise measurement of the relative positions of the chromatic cores $\Delta{z_{c}}(\nu_{1},\nu_{2}) = z_{c}(\nu_{1}) - z_{c}(\nu_{2})$, where $z_{c}$ is the distance of the core from the black hole and $\nu_{2} > \nu_{1}$. For this reason, we analyzed the core-shift in the 1928+738 jet before proceeding to the analysis of jet geometry and kinematics.

\begin{table}[t!]
\caption{Core-Shift measurements for each pair of frequencies. \label{table:core_shift_full_pair}}
\centering
\begin{tabular}{ccccc}
\hline
\hline
$\nu_{2}$ (GHz) & $\nu_{1}$ (GHz) & $\Delta z_{\rm core}$ (mas) & Angle ($^\circ$) & $\rho_{\rm max}$ \\ \hline

\multicolumn{5}{c}{Adjacent Frequency Pairs} \\\hline
23.88 & 15.15  & 0.136 $\pm$ 0.033 & 180.0 & 0.984\\
15.15 & 7.65   & 0.247 $\pm$ 0.067 & 166.0 & 0.999\\
7.65  & 4.15   & 0.201 $\pm$ 0.120 & 153.4 & 0.999\\
4.15  & 1.55   & 0.923 $\pm$ 0.314 & 162.3 & 0.997\\ \hline
\multicolumn{5}{c}{Other Frequency Pairs} \\\hline
23.88 & 7.65   & 0.361 $\pm$ 0.067 & 175.2 & 0.996\\
23.88 & 4.15   & 0.489 $\pm$ 0.120 & 169.4 & 0.989\\ 
23.88 & 1.55   & --- & --- & ---\\
15.15 & 4.15   & 0.437 $\pm$ 0.120 & 164.1 & 0.995\\
15.15 & 1.55   & --- & --- & ---\\
7.65 & 1.55    & 0.955 $\pm$ 0.314 & 174.0 & 0.971\\ \hline
\end{tabular}
\tablefoot{Columns from \emph{left to right}: (1), (2) observing frequencies; (3) magnitude of the core-shift; (4) direction of the core-shift; (5) maximum cross-correlation coefficient.}
\end{table}

We used the ``self-referencing'' method, which is based on a 2D cross-correlation of optically thin jet structures a few mas away from the core \citep[see][for details]{Croke2008, O'Sullivan2009, Sokolovsky2011, Pushkarev2012, Fromm2013, Fromm2015, Pushkarev2019, Park2021}.

To enable an accurate comparison, we reconstructed the jet images at 1 -- 24\,GHz listed in Table~\ref{table:VLBA} (see also, Figure~\ref{Fig:imgs}, (a) -- (e)), which are obtained simultaneously. For each frequency pair, the images are adjusted to have the same \emph{uv} range and are convolved with the same circular beam. Likewise, the pixel sizes are chosen to be identical. From the subsequent cross-correlation, we determined the core-shift vector for the given frequency pair. We follow the convention \citep[e.g.,][]{Fromm2013, Kutkin2014}, to assume the uncertainties of the core-shift magnitude to be 1\,/\,20 of the convolved beam. We list all core-shifts thus derived in Table\,\ref{table:core_shift_full_pair}.

\begin{figure}[t!]
\centering
\includegraphics[width=0.47\textwidth]{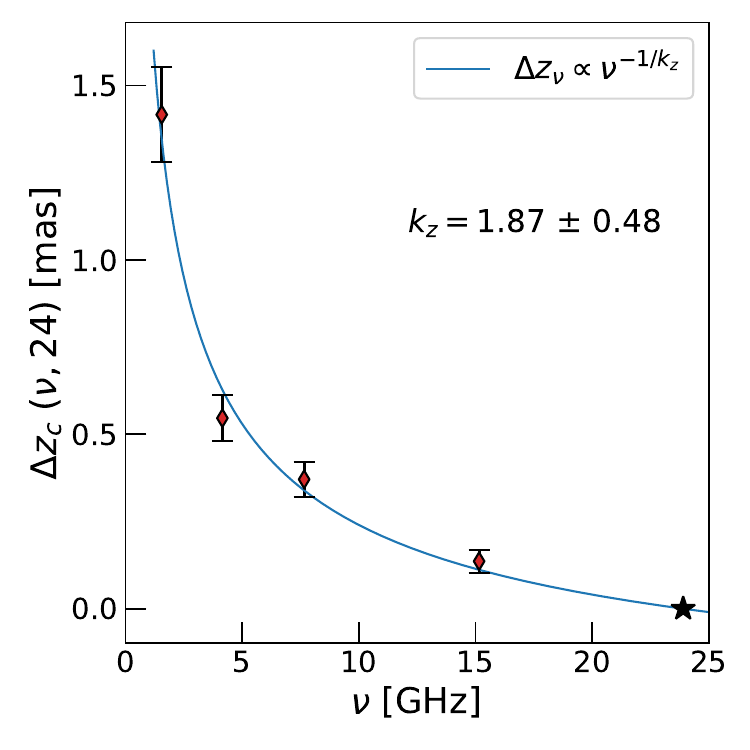}
    \caption{Relative core positions of the 1928+738 jet as function of frequency, with respect to the 24\,GHz core. The solid line shows the best-fit power-law function $\Delta{z_{c}} = \Omega\,(\nu^{-1/k_{z}}_{\rm GHz} - 24^{-1/k_{z}})$.}
    \label{Fig:core-shift}
\end{figure}

At each frequency we computed the core position with respect to the one at the highest frequency. The relative positions are presented in Figure\,\ref{Fig:core-shift}. A systematic decrease with increasing frequency is obvious. We describe the data with a quasi-reciprocal function, $z_{c}\,\propto\,\nu^{-1/k_{z}}$ \citep{Blandford1979, Konigl1981}, where the value of $k_{z}$ is determined by the magnetic field and electron density distribution along the jet. The special case $k_{z}\,=\,1$ corresponds to the case of synchrotron self-absorption dominated nuclear opacity in a conical jet with energy equipartition. We performed a least-square fit using the function $\Delta{z_{c}} (\nu, 24)\,=\,\Omega\,(\nu^{-1/k_{z}}_{\rm GHz} - 24^{-1/k_{z}})$, where $\Delta{z_{c}} (\nu, 24)$ is the core position at frequency $\nu$ with respect to the one at 24\,GHz. The best-fit parameter values are $\Omega=2.21\pm0.15$ and $k_{z}=1.87\pm0.48$ (see Figure\,\ref{Fig:core-shift}). The best-fit shows a reduced chi-square value $\chi^2/{d.o.f} = 1.38$; within the errors, the best-fit value of $k_{z}$ is marginally consistent with unity. An extrapolation of the best-fit line to higher frequencies is highly uncertain. Rather, we assume that the core-shift of the 24\,GHz core is negligible, $\Delta{z_{c}} (\nu, 24)\,\approx\,0$, even for $\nu \rightarrow \infty$.

All data for the 1928+738 jet presented in this paper are corrected for core-shift. A deeper analysis of the core-shift will be presented in a forthcoming paper, with a wider frequency coverage and a denser sampling of observing frequency.

\subsection{Possible limb brightening \label{subsec:limb}}

\begin{figure*}[ht!]
\centering
\includegraphics[width=\textwidth]{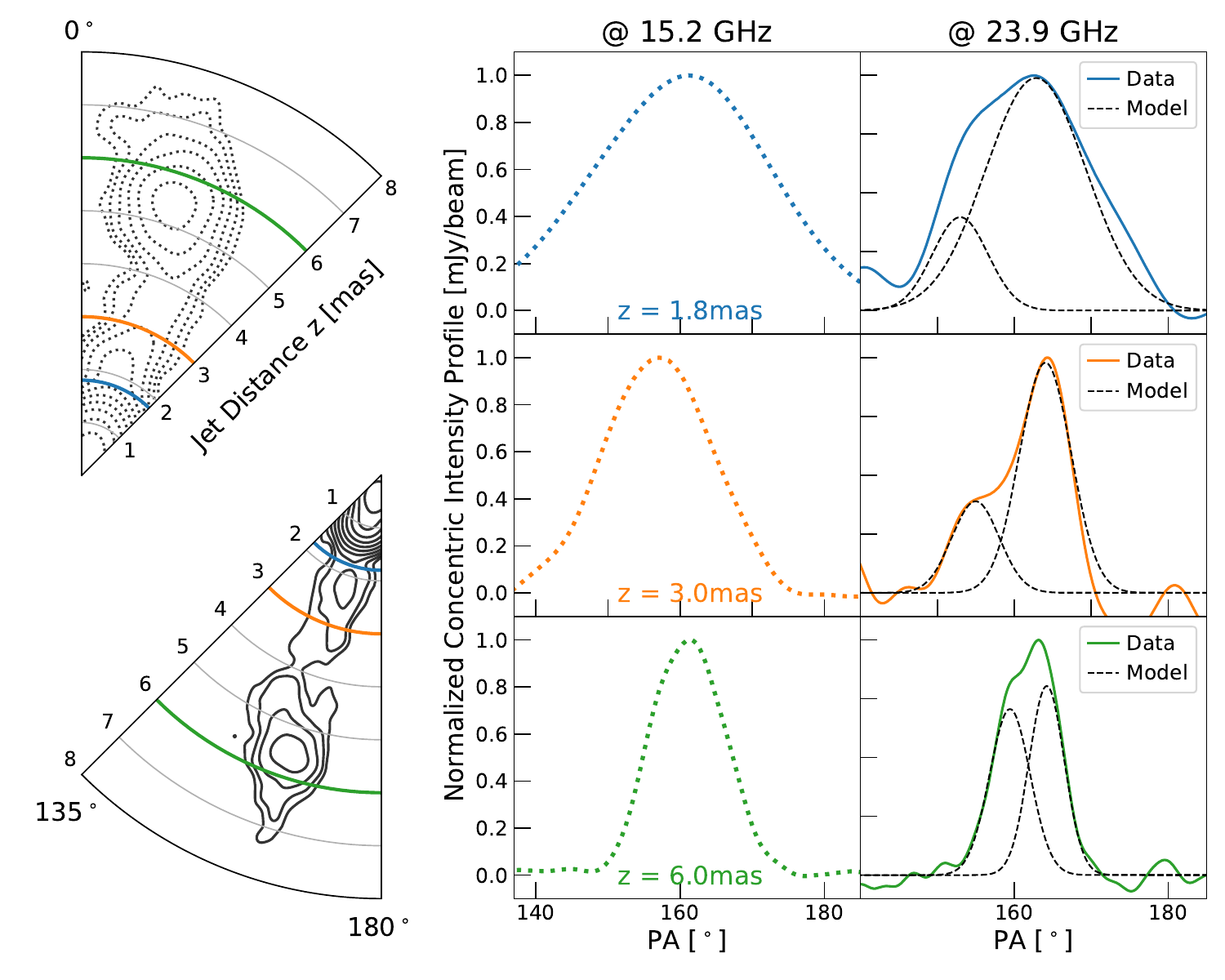}
    \caption{ \emph{\textbf{Left:}} Two of our VLBA maps in polar coordinates, convolved with circular beams. The black solid and gray dashed contours represent the 24\,GHz and 15\,GHz images, respectively. The 15\,GHz image is rotated by 180$^\circ$ to ease the presentation. \emph{\textbf{Right:}} Three concentric jet intensity profiles obtained at different angular distances $z$ from the origin of the jet (represented by different colors). The solid and dotted lines denote the intensities at 24\,GHz and 15\,GHz, respectively. All profiles are normalized to unit maximum.}
    \label{Fig_limb}
\end{figure*}

We restored the jet images using circular beams with diameters corresponding to the geometric mean of major and minor axis of the synthesized beams. This image reconstruction step removes distortion related to the beam shapes. A constant P.A. is usually adopted to describe the direction of a jet \citep[e.g.,][]{Nakahara2018, Hada2018, Nakahara2019, Nakahara2020, Park2021}; however, the jet of 1928+738 is curved. We therefore defined a jet axis by determining the locations of intensity maxima in the image plane \citep[see][]{Pushkarev2017, Kovalev2020, Okino2022}: Using polar coordinates with the core as the center point, we measure the jet intensity profile along the azimuthal direction at multiple radii. At first glance, the jet appears to have a single ridge; however, we found that the concentric intensity profile of the jet can only be described with two Gaussian functions in several cases, especially at 24\,GHz. In Figure\,\ref{Fig_limb}, we present three examples of such profiles. We note that limb brightening in 1928+738 has not been reported before, probably due to the lower angular resolution and limited sensitivity of previous observations \citep[e.g., VLBA 15\,GHz image of][]{Pushkarev2017}. We plan to explore the limb brightening structure in 1928+738 jet with future observations with higher angular resolution and higher sensitivity (K. Yi et al. 2024, in preparation).

\subsection{Jet Width Profile \label{subsec:jetwidth}}

The jet radius $R$ (i.e., the half width) scales with distance from the black hole, $z$, like $R\,\propto\,z^{m}$. The power-law index $m$ is close to unity in case of the classical conical jet, while $m\,<\,1$ is expected for collimated jets \citep[e.g.,][]{Asada2012,Hada2018,Park2021}. The images of 1928+738 (Figure\,\ref{Fig:imgs}) allow us to analyze the radius of its jet over a wide range of spatial scales.

\begin{figure}[t!]
\centering
\includegraphics[width=0.47\textwidth]{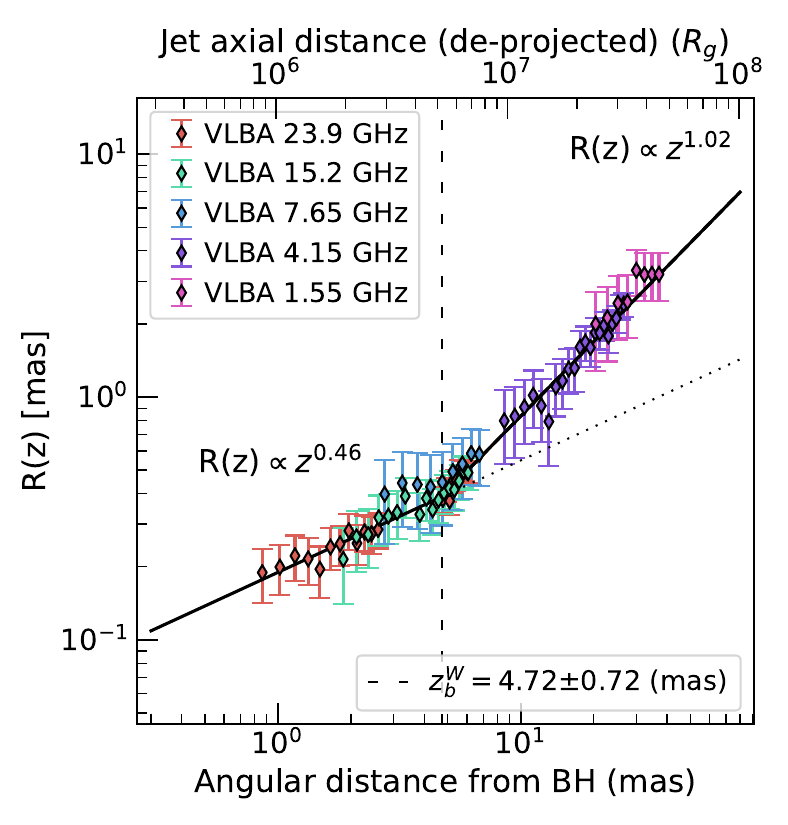}
    \caption{Jet radius profile as a function of distance $z$ from the black hole. The distance is displayed in units of $R_{g}$ (top axis) and mas (bottom axis). Error bars represent $\pm\,1\,\sigma$ uncertainties. The solid line represents the best-fit broken power-law model. The dotted line represents an extrapolation of the inner part of the best-fit model. The vertical dashed line indicates the location of the break in the jet width profile, $z_{b}^{w}$. 
}
    \label{Fig:CZ}
\end{figure}

We obtained transverse slices along the jet in steps of 1\,/\,3 of the restoring beam size. We measured the full width half maximum (FWHM) of each slice and subtracted the beam FWHM in squares (deconvolution). The FWHM is measured by fitting the intensity profile with a Gaussian function. Whenever a single Gaussian function can not describe the profile well, a double Gaussian is used. In those cases, we use the distance between the outer half-maximum points as a measure of the jet width \citep{Hada2013, Hada2016, Park2021}. We considered the measurements to be valid only when (i) adjacent distance bins are smoothly connected, and (ii) the peak intensity of the slice exceeds 14 times the image rms level. The first few measurements, starting at the center point, are strongly affected by the core brightness; we therefore discard the distance bins out to 1.5 times of the beam size away from the core. In this paper, we define the jet radius $R$ as 1\,/\,2 of the deconvolved FWHM and use it as an indicator of the jet width. The uncertainties of the jet radii are assumed to be 1\,/\,10 of the beam sizes \citep[see][for details]{Park2021}. In this way, we could derive the jet radii covering distances from $\sim0.8$\,mas to $\sim40$\,mas. We present the jet radius as a function of distance from the black hole in Figure\,\ref{Fig:CZ}. In general, the jet radii from different frequencies and different angular resolutions are consistent with each other within errors (1\,$\sigma$ confidence intervals) at given distance bins. 

We modelled the jet radius as function of distance with a broken power-law, meaning in log space
\begin{eqnarray}
    \rm ln \ R = \left\{ \begin{array}{ll}
    {m_{1}z + b } , & \mbox{ $z < z_{b}^{w}$} \\
    {m_{2}z + b + (m_{1}-m_{2})\,z_{b}^{w}} , & \mbox{ $z > z_{b}^{w}$} \\\end{array} \right.  
\label{eq:jet_radius}
\end{eqnarray}
where $z_{b}^{w}$ is the location of the break point, $m_{1,2}$ are the slopes, and $b$ is a constant offset. This function describes the jet profile well over the full range of distances from the black hole. The best-fit profile has $z_{b}^{w}$ = (4.72 $\pm$ 0.72) mas, with $R \propto z^{0.46 \pm 0.10}$ (i.e., parabolic shape) and $R \propto z^{1.02 \pm 0.03}$ (i.e., conical shape), before and after the break point, respectively.

Our fit is based on the assumption that the input errors solely depend on the beam size. Although our broken power-law model apparently fits the data well (see the solid line in Figure\,\ref{Fig:CZ}), the goodness of the fit should be compared with a single power-law model quantitatively. We used the Bayesian information criterion (BIC), defined as ${\rm BIC} \equiv -2 \ln \mathcal{L}_{\rm max} + k \ln {N}$. Here, $\mathcal{L}_{\rm max}$ is the maximum likelihood of a model, $k$ and ${N}$ are the number of free parameters and the number of data points, respectively. Adopting the standard assumption of Gaussian errors, ${\rm BIC} = \chi^2 + k \ln {N}$. The BIC of our best-fit broken power-law model is smaller than the BIC of the best-fit single-power law model by $\rm \Delta{BIC}\approx 34$. Accordingly, a broken power-law model is strongly favored statistically, indicating a robust discovery of the jet collimation break (JCB) in FSRQ 1928+738.

\subsection{Jet Velocity Field \label{subsec:velfield}}  

\begin{figure}[t!]
\centering
\includegraphics[width=0.47\textwidth]{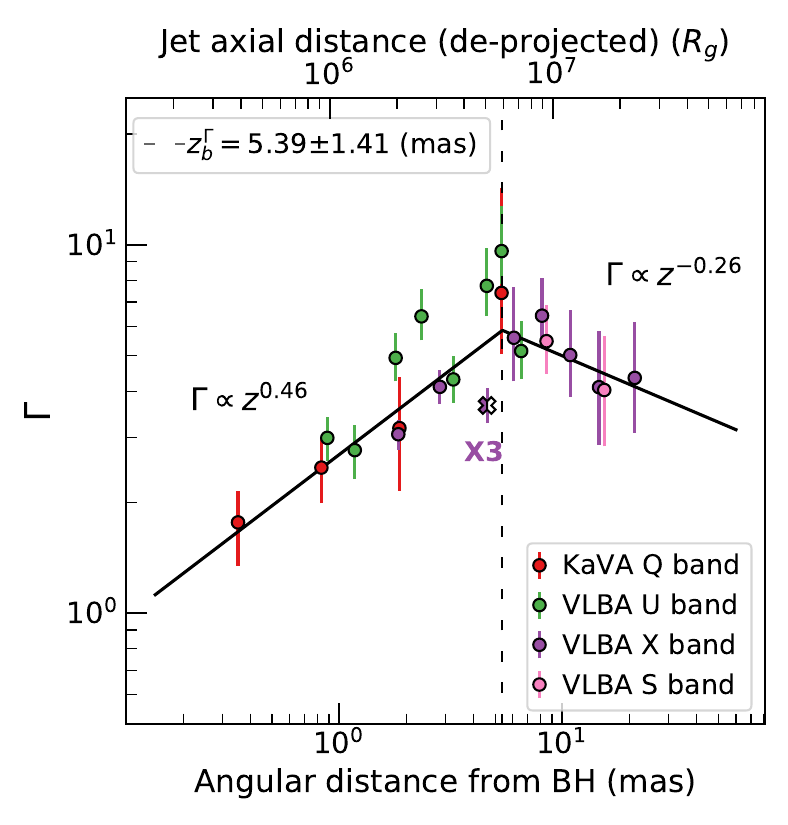}
    \caption{The bulk Lorentz factor as a function of distance $z$ from the black hole. The distance is displayed in units of $R_{g}$ (top axis) and mas (bottom axis). Colors indicate observations at different bands from various monitoring programs. Error bars represent $\pm\,1\,\sigma$ uncertainties. The solid line represents the best-fit broken power-law model. One of the knots traced at X band (X3), marked by an empty colored cross, is an outlier from the general relation. The vertical dashed line indicates the location of the transition point of the jet kinematics, $z_{b}^{\Gamma}$.
}
    \label{Fig:AZ}
\end{figure}

We constrain the jet velocity field of FSRQ 1928+738 over the entire region inside and outside the JCB. We measured the proper motions of distinctive structures. As previously reported \cite[e.g.,][]{Eckart1985}, the brightness distribution of the jet is modeled well with a set of circular Gaussian components at all frequencies (see Section\,\ref{subsec:coreshift}). We hereafter refer to the \emph{modelfit} components as ``knots''. For our kinematic analysis we cross-identify and trace knots across different epochs. We used multi-epoch monitoring data from the KaVA Q band, MOJAVE U band, and the Astrogeo X and S band, respectively (see Section\,\ref{sec:obs}). We note that we did not attempt cross-band identification of knots. The proper motion $\mu_{z}$ (in mas/yr), and its corresponding apparent speed $\beta_{\rm app} (\equiv v_{\rm app}/c)$ of the each knot feature were derived by fitting their radial distance $z$ with time. We detected proper motions at distances between $\sim0.3$ and $\sim20$~mas from the black hole. We provide the details for our \emph{modelfit} analysis in Appendix\,\ref{sec:appendix_kin}. 

We assume that the observed proper motions reflect the bulk motion of the jet flow \citep[e.g.,][]{Lister2009, Nakamura2013}, given the continuous acceleration and deceleration profiles observed in various data across different frequencies and observing periods. However, it is noteworthy that the observed speeds, based on \emph{modelfit} analysis, may not always correspond to the bulk jet speeds; instead, they may represent pattern speeds caused by, for example, jet instabilities \citep[e.g.,][]{Hardee2000} or stationary shocks.

\begin{figure*}[ht!]
\centering
\includegraphics[width=\textwidth]{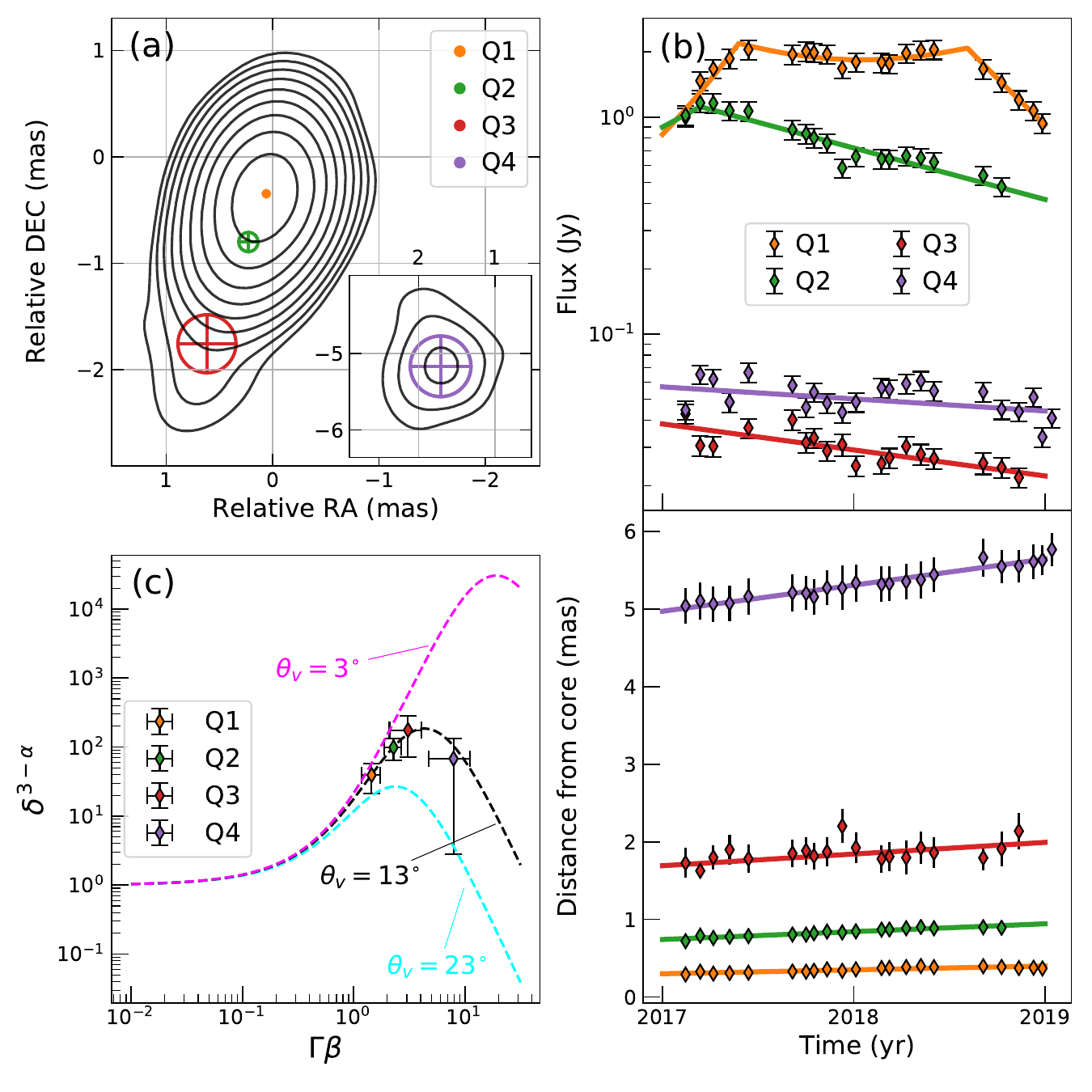}
    \caption{(a) Zoom-in view on the upstream region of the KaVA 43\,GHz image in Figure~\ref{Fig:imgs}. The isolated jet feature at $\approx5$\,mas is embedded in the bottom right of the plot. The four knots, identified via \emph{modelfit} analysis, are marked in the map. (b) The flux density and the displacements of the knots as functions of time. Exponential fits to the flux density data and linear fits to the motions are shown as continuous lines. (c) The relativistic beaming factor $\delta^{3-\alpha}$, as a function of $\Gamma\beta$, assuming a spectral index of $\alpha=-0.5$. We show our data along three theoretical lines corresponding to different values of $\theta_{\rm v}$.
    }
    \label{Fig_KaVA_var}
    
\end{figure*}

We detected significant outward motions for most (22 out of 25) of the knots we traced. The proper motion of three knots is not significantly different from zero, so we do not include them into our analysis. We then converted the apparent speed $\beta_{\rm app}$ of each knot into the intrinsic speed $\beta = \beta_{\rm app}\,/\,(\sin{\theta_{\rm v}} + \beta_{\rm app}\cos{\theta_{\rm v})}$ and derived the bulk Lorentz factor $\Gamma = 1\,/\,\sqrt{(1-\beta^2)}$ by adopting a viewing angle of $\theta_{\rm v}\,=\,13.2^{\circ}$ (see Section\,\ref{subsec:KaVA_var}). In Figure~\,\ref{Fig:AZ}, we present the radial profile of the bulk Lorentz factor. Interestingly, a systematic increase and a subsequent decrease of $\Gamma$ are clearly visible in the velocity field, with a peak approximately at 5\,mas. This indicates that the bulk jet is gradually accelerated in the inner region within a distance of $\approx5$\,mas, and decelerated at larger distances. The peak $\Gamma_{\rm max} \approx 10$, calculated from the maximum observed speed $\beta_{\rm app} = 7.2$, is broadly consistent with previous measurements by \citet{Lister2019} and \citet{Kun2014}. We fit a broken-power law to the $\Gamma$ profile, in the same way as we did for the jet radius $R(z)$. During the fitting process we noted that X3, one of the X band knots, is an outlier which may not be associated with the underlying bulk flows, and we excluded it from the fit (this is discussed in Appendix\,\ref{sec:kin_outlier}). We found a clear division into an inner and an outer velocity field, with $\Gamma \propto z^{0.46\pm0.08}$ and $\Gamma \propto z^{-0.26\pm0.25}$, connected at the break point $z_{b}^{\Gamma}=(5.39\pm1.41)$\,mas.

\begin{table*}[ht!]
\centering
\caption{Kinematics and Physical Parameters of the KaVA 43\,GHz Jet Features
 \label{table:KaVA_params}}
\begin{tabular}{c c c c c c c c c}
\hline
\hline
Knot ID & $\langle \rm z \rangle$ (mas) & $a$ (mas) & $\rm \tau_{dec}$ (yr) & $\delta$ & $\beta_{\rm app}$ & $\Gamma_{\rm var}$ & $\theta_{\rm v}$ ($^\circ$) \\ \hline
Q1$^{*}$ & 0.35 & 0.064 & 0.68 $\pm$\ 0.09 & 2.85 $\pm$\ 0.38 & 0.95 $\pm$\ 0.51 & 1.76 $\pm$\ 0.22 & 13.29 $\pm$\ 6.39 \\
Q2 & 0.83 & 0.224 & 1.83 $\pm$\ 0.19 & 3.72 $\pm$\ 0.38 & 1.92 $\pm$\ 0.67 & 2.49 $\pm$\ 0.37 & 13.11 $\pm$\ 3.27 \\
Q3 & 1.86 & 0.522 & 3.62 $\pm$\ 0.62 & 4.38 $\pm$\ 0.75 & 2.86 $\pm$\ 1.40 & 3.24 $\pm$\ 0.93 & 12.24 $\pm$\ 3.69 \\
Q4 & 5.34 & 0.858 & 7.82 $\pm$\ 2.14 & 3.34 $\pm$\ 0.91 & 6.40 $\pm$\ 1.50 & 7.96 $\pm$\ 3.15 & 14.06 $\pm$\ 2.50 \\ \hline

\end{tabular}
\tablefoot{$^{*}$ The flux decay timescale of Q1 is derived by adopting the empirical relation $\tau_{\rm dec}\,=\,1.3\,\times\,\tau_{\rm rise}$ \citep{Valtaoja1999}. The other physical parameters $\delta$, $\Gamma_{\rm var}$ and $\theta_{\rm v}$ are computed from it.}\\

\end{table*}

Due to the limited range of distances in the jet of 1928+738 (\textless\,1 order of magnitude), the outer velocity field is poorly constrained. Even though, the best fit describes the $\Gamma$ profile well, with $\chi^2/{d.o.f}\,=\,1.15$. Moreover, we note that the outer trend of jet deceleration is in a good agreement with the slow jet speed on arcsecond (and kpc) scales, which can be inferred from VLA\,/\,MERLIN observations (see Appendix\,\ref{sec:cj_speed}). We can therefore conclude that we discovered a distinct transition in the jet acceleration profile, a jet acceleration break (JAB), in FSRQ 1928+738. 

The discovery of the jet acceleration and deceleration also implies the large dynamic range in the velocity of 1928+738 jet. However, the effectiveness of $\beta$ and $\Gamma$ in demonstrating the jet speed is limited to the non-relativistic regime ($v\,\ll\,c$) and the relativistic regime ($v\,\rightarrow\,c$) respectively. Thus we will hereafter use $\Gamma\beta$ as a proxy for the bulk speed, which can represent both regimes.

\subsection{Doppler factor and viewing angle \label{subsec:KaVA_var}}

The motion of jet components toward the observer at relativistic speeds results in a boosting of the observed flux densities, and a reduction of observed timescales. Beaming effects scale with the Doppler factor $\delta \equiv [\Gamma (1\,-\,\beta\cos{\theta_{\rm v}})]^{-1}$. In this section, we analyze our KaVA 43\,GHz monitoring data, and estimate the variability Doppler factor for 1928+738.

In our KaVA 43\,GHz images, we cross-identified four knots, labeled Q1, Q2, Q3 and Q4 in order of increasing distance from the core. Their locations in a total intensity map and their light-curves are presented in Figure~\ref{Fig_KaVA_var}. Thanks to its high cadence, our KaVA monitoring enables us to use an approach suggested by \citet{Jorstad2005, Jorstad2017} for deriving the Doppler factor. This approach relies on the evolution of \emph{modelfit} parameters (i.e., flux, position, and size) of knots. We also refer the reader to \citet{Weaver2022}.

We estimated the uncertainties of \emph{modelfit} parameter values following the methods suggested by \citet{Lister2009} and \citet{Fomalont1999}. We first considered a typical calibration error of 10\,--\,15\% in the flux density. The positional uncertainty is set to one fifth of the beam size, with the ratio of beam size and peak-to-noise ratio added in quadrature. In case of very compact and bright knots, these uncertainties are reduced by a factor of 2. For the uncertainty in angular size, $\sigma_{a}$, we adopted the empirical scaling relation between size and brightness temperature $T_{b}$ by \citet{Jorstad2017}:

\begin{equation}
\left\{ \begin{array}{ll}
    {T_{b} = 7.5 \times 10^{8} \ S\,/\,a^2}  \\
    {\sigma_{a} \approx 6.5 \times T_{b}^{-0.25}} \\\end{array} \right.
\label{eq:Tb}
\end{equation}
where $T_{b}$ is in units of K, the flux $S$ is in units of Jy, and the size $a$ is in units of mas.

We modeled the light curve of each knot as a sequence of an exponential rise and a subsequent exponential decay \citep{Valtaoja1999, Hovatta2009}:
\begin{eqnarray}
    {\rm ln} \ (S(t)\,/\,S_{\rm o}) = \left\{ \begin{array}{ll}
    {(t-t_{\rm max})\,/\,\tau_{\rm rise}} , & \mbox{ $t < t_{\rm max}$} \\
    {-(t-t_{\rm max})\,/\,\tau_{\rm dec}} , & \mbox{ $t > t_{\rm max}$} \\\end{array} \right.  
\label{eq:lightcurve}
\end{eqnarray}
where $S_{o}$ is the peak flux, $t_{\rm max}$ is the time of peak flux, and $\tau_{\rm rise}$ and $\tau_{\rm dec}$ denote the rise and decay timescales, respectively. We found that the brightest knot Q1 experienced exponential flaring (i.e., rise and decay) twice during the observing period. One complete flare is also found in the second brightest knot, Q2. The remaining two knots only show exponential decays. The exponential fits to the light-curves are illustrated in Figure~\ref{Fig_KaVA_var}b. The observed timescales are related to the intrinsic ones through  $\tau_{\rm obs} \equiv \tau_{\rm int}\,/\,\delta$. We identify the variability timescale $\tau_{\rm dec}$ of a given knot with the light-travel time across the knot, which is valid as long as the flux decay is caused by radiative cooling. Under this assumption, the variability Doppler factor $\delta$ can be expressed as 

\begin{equation}\label{eq4}
        \delta = 25.3\ \frac{(a \ [{\rm mas^{-1}}]) \ (D_{\rm L} \ [{\rm Gpc^{-1}}]) }{(\tau_{\rm dec} \ [{\rm yr^{-1}}]) \ (1+z)}
\end{equation}
where $a$ is knot size (FWHM) and $D_{\rm L}$ is the luminosity distance. This approach allows us to compute the Doppler factors of the four individual knots. In addition, Eq.~\ref{eq4} is free from commonly used, but highly idealized, assumptions such as energy equipartition between the radiating particles and the magnetic field, or the viewing angle satisfying $\sin{\theta_{c}}=1\,/\,\Gamma$. For our calculations, we used the knot sizes measured at times of maximum knot brightness \citep{Casadio2015, Casadio2019}.

By combining the variability Doppler factor $\delta$ and the apparent speed $\beta_{\rm app}$ of each knot, we can infer the variability Lorentz factor $\Gamma_{\rm var}$ and viewing angle $\theta_{\rm v}$ from 

\begin{equation}\label{eq5}
        \Gamma_{\rm var} = \frac{\beta^{2}_{\rm app} + \delta^{2} + 1}{2 \delta}; ~~ 
        \theta_{\rm v} = {\rm arctan}\frac{2\beta_{\rm app}}{\beta^{2}_{\rm app} + \delta^{2} - 1} .
\end{equation}
The results of our KaVA 43\,GHz analysis are summarized in Table~\ref{table:KaVA_params} which shows the values of the mean radial distance $\langle z \rangle$, the angular size $a$, the flux decaying timescale $\tau_{\rm dec}$, the Doppler factor $\delta$, the Lorentz factor $\Gamma_{\rm var}$, and the jet viewing angle $\theta_{\rm v}$ derived for each of the four knots. The uncertainties of $\delta$, $\Gamma_{\rm var}$ and $\theta_{\rm v}$ are estimated by means of standard error propagation.

Unfortunately, the analysis of knot Q1 appears rather unreliable. The exponential fits to the light-curve do not constrain the decay timescales in a meaningful way, resulting in $\tau_{\rm dec}=(0.98\pm1.81)$\,yr and $\tau_{\rm dec}=(0.42\pm0.60)$\,yr for the first and second flare respectively, meaning we have to disregard the fit results for our analysis \citep[see e.g.,][]{Weaver2022}. Instead, we apply the empirical relation $\tau_{\rm dec}\,=\,1.3\,\times\,\tau_{\rm rise}$ \citep{Valtaoja1999}, which provides $\tau_{\rm dec} = 0.68\pm0.09$\,yr and leads to the values for $\delta$, $\Gamma_{\rm var}$ and $\theta_{\rm v}$ listed in Table\,\ref{table:KaVA_params}.

The 1928+738 jet is characterized by mild variability, which indicates a relatively large viewing angle. Notably, we found an initial increase and subsequent decrease of the Doppler factor $\delta$ along the knots from Q1 to Q4, which presumably resulted from jet acceleration. This implies a specific brightness evolution, since relativistic beaming scales with a factor $\delta^{3-\alpha}$, where $\alpha$ is the spectral index defined by $I_{\nu}\,\propto\,\nu^{\alpha}$. In Figure\,\ref{Fig_KaVA_var}, the beaming factor for the case $\alpha=-0.5$ is displayed as a function of $\Gamma\beta$. The data points for the four knots match the theoretical line for $\theta_{\rm v}=13^{\circ}$. Moreover, we found that the viewing angle $\theta_{\rm v}$ remains constant along the jet within errors. Indeed, our measurements provide observational evidence for the first time that jet collimation or acceleration (described in Section\,\ref{subsec:jetwidth} and \ref{subsec:velfield}) is unrelated to changes in the viewing angle. We adopt the average value $\theta_{\rm v}\,=\,13.2^{\circ}\,\pm\,2.1^{\circ}$ throughout this paper.

\section{Discussion}
\label{sec:discussion}

\begin{figure}[t!]
\centering
\includegraphics[width=0.47\textwidth]{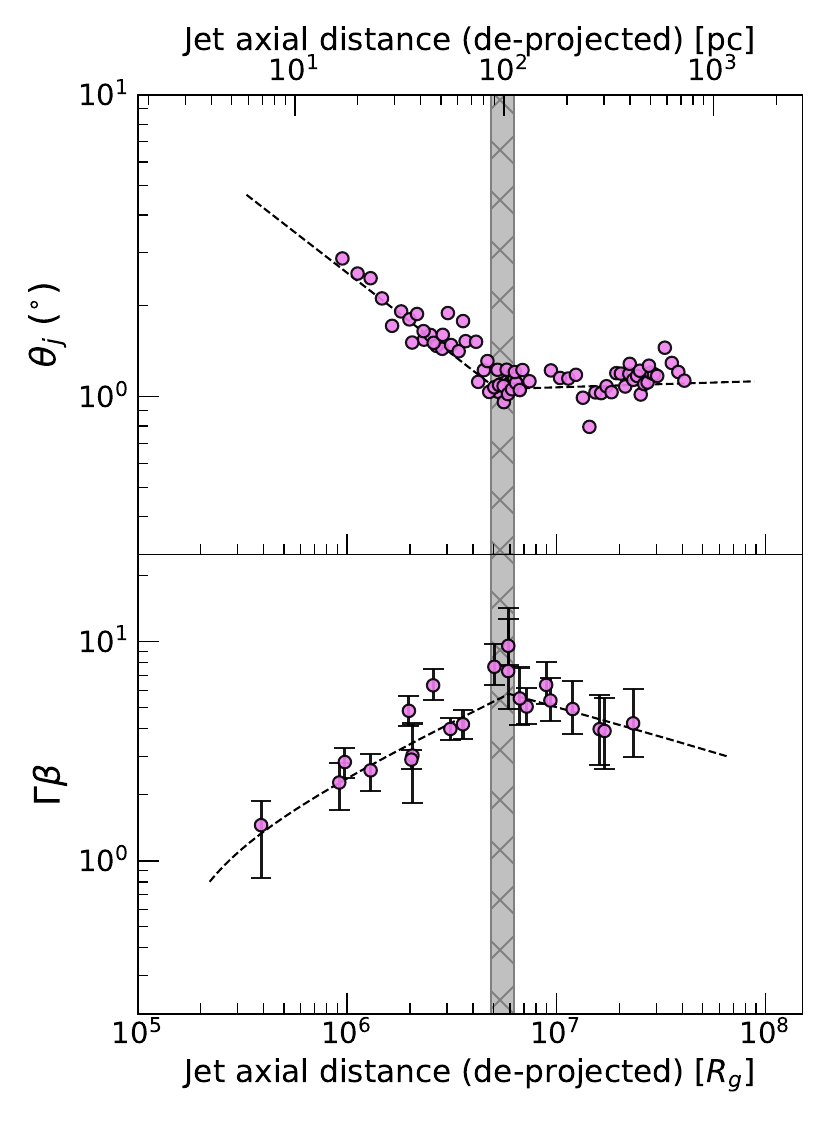}
    \caption{Collimation and acceleration of 1928+738 jet as a function of distance from the black hole. The distance is displayed in units of parsec (top axis) and $R_{g}$ (bottom axis). \emph{Top:} Intrinsic half opening angle $\theta_{j}$ as derived from the observed jet width profile. \emph{Bottom:} $\Gamma\beta$ calculated from the proper motions. The shaded area denotes the location where the ACZ terminates, $\langle z_{b} \rangle = (5.6\,\pm\,0.7) \times 10^6\,R_{g}$.
}
    \label{Fig_coex}
\end{figure}

In Figure\,\ref{Fig_coex}, we visualize the jet collimation and acceleration of 1928+738, by displaying the deprojected half opening angle $\theta_{j}$ and $\Gamma\beta$ as function of deprojected distance from the black hole in units of $R_{g}$ and parsec. The half-opening angle declines down to $\approx 1^{\circ}$ at around 100\,pc. Meanwhile, $\Gamma\beta$, which is assumed to represent the bulk jet speed, increases with distance and reaches its maximum $\Gamma\beta \sim 10$, likewise at $\sim$\,100\,pc. Evidently, jet acceleration and collimation occur in the same region, the ACZ. Beyond the ACZ, one may expect the jet to decelerate slowly, with the opening angle $\theta_{j}$ remaining constant. Figure\,\ref{Fig_coex} shows that the data are consistent with this expectation. We identified the downstream end of the ACZ by independently measuring the locations of the jet collimation and acceleration breaks $z_{b}^{w}$ and $z_{b}^{\Gamma}$ (see Section\,\ref{subsec:jetwidth} and \ref{subsec:velfield}). The two values are in good agreement; the shaded area in Figure\,\ref{Fig_coex} denotes their average and the corresponding 1$\sigma$ confidence interval, $\langle z_{b} \rangle = (5.6 \pm 0.7) \times 10^6\,R_{g}$. In this section, we will discuss the jet collimation and jet acceleration of 1928+738 one by one.

\subsection{Jet Collimation \label{subsection:soi}}

A number of analytical and numerical studies \citep[e.g.,][]{Komissarov2007, Lyubarsky2009, Nakamura2018} have shown that external pressure $p_{\rm ext}$ is responsible for jet collimation. The external pressure profile $p_{\rm ext} \propto z^{-\kappa}$ needs to be sufficiently flat (i.e., $\kappa\leqq2$) to collimate the jet \citep[e.g.,][]{Komissarov2009}. The source of $p_{\rm ext}$ is most likely the ambient medium confining the jet which is captured by the gravitational field of the black hole. Thus, the range of gravitational influence of the black hole in the centre of 1928+738 is of particular interest. 

The sphere of gravitational influence (SOI) can be defined as $R_{\rm SOI} \equiv GM_{\bullet}/ \sigma_{*}^2$, where $\sigma_{*}$ is the stellar velocity dispersion \citep{Peebles1972} in the center of the host galaxy. Assuming that the stars and the ionized gas around the black hole (e.g., [O\,{\footnotesize III}]$\lambda 5007$ line-emitting gas) are bound by the same gravitational potential \citep[e.g.,][]{Nelson1996, Boroson2003}, $\sigma_{\rm [O\,{\footnotesize III}]}$ can be used as a proxy for $\sigma_{*}$. We adopted $\sigma_{\rm [O\,{\footnotesize III}]}$ = 166\,km/s for 1928+738 \citep{Marziani1996, Bian2004}, which gives us $R_{\rm SOI} \approx 3.3 \times 10^{6}\,R_{g}$. However, one may expect $\sigma_{\rm [O\,{\footnotesize III}]}\,>\,\sigma_{*}$, due to the contribution of non-gravitational motion (like turbulence) to the gas dynamics \citep[e.g.,][]{Greene2005, Woo2006, Bennert2018, Sexton2021}. Overall, our value for $R_{\rm SOI}$ can be considered a rough estimate and/or a lower limit. An additional estimate for $\sigma_{*}$ is provided by the empirical $M_{\bullet}-\sigma_{*}$ relation \citep[e.g.,][]{Kormendy2013}. In this case, $R_{\rm SOI}$ is solely determined by the black hole mass; for 1928+738, $M_{\bullet} \approx 3.7 \times 10^{8}\,M_{\odot}$ and thus $R_{\rm SOI} \approx 2.1 \times 10^{6}\,R_{g}$.

Both estimates of $R_{\rm SOI}$ show an order-of-magnitude consistency with $\langle z_{b} \rangle$, the location where the ACZ terminates in 1928+738. This is the first time that such a coincidence has been reported for a quasar \citep{Okino2022, Burd2022}, although such a scaling has been found in a few nearby radio galaxies. \citet{Asada2012} suggested that the JCB in M\,87 occurs near the sphere of influence, which is characterized by the Bondi radius $R_{B} \equiv 2GM_{\bullet}/ c_{s}^2$ (where $c_{s}$ is the sound speed of X-ray emitting hot gas). Likewise, in NGC\,6251 the JCB is found at approximately $R_{\rm SOI}$ \citep{Tseng2016}. The remarkable similarity of the situation in 1928+738 with the one in low-luminosity AGNs suggests that an interplay between the jet and external medium could be a fundamental requirement for AGN jet collimation. In the case of low-luminosity AGNs, non-relativistic winds from the hot accretion flow \citep[e.g.,][]{Yuan2015} within the Bondi radius\,/\,SOI have been suggested to serve as the external confining medium \citep[e.g.,][]{Nakamura2018, Park2019-Fara}. Likewise, such winds are a likely candidate for medium surrounding the jet in 1928+738. If so, it is necessary to investigate the wind properties within the ACZ of this quasar, for which a cold\footnote{Cold accretion flows are presumed to be characteristic for accretion at relatively high Eddington ratio $\dot{M}/\dot{M}_{\rm Edd} \gtrsim 1\%$ \citep[e.g.,][]{Ghisellini2011, Heckman2014}. We note that the Eddington ratio of 1928+738 has been found to be $\gtrsim60\%$ \citep[e.g.,][]{Park2017}.} accretion flow is expected \citep[e.g.,][]{Crenshaw2003}. Future VLBI polarimetric observations for Faraday rotation may provide more information about the ambient medium (K. Yi et al. 2024, in preparation).

\subsection{Jet Acceleration \label{subsection:magnetic nozzle}}

In Figure\,\ref{Fig_coex}, we see that the 1928+738 jet is gradually accelerated to relativistic speeds over a range of several million $R_{g}$, which is consistent with the basic characteristics of MHD acceleration \citep[e.g.,][]{Vlahakis2004, Beskin2010}. The jet acceleration is accompanied by jet collimation, as the MHD jet acceleration model predicts. Such models also predict that collimation and acceleration of a jet stop at the same distance from the black hole -- which is exactly what we observe in 1928+738, and what has been observed in M\,87 \citep[e.g.,][]{Asada2012, Asada2014}, NGC\,315 \citep{Park2021}, and 1H\,0323+342 \citep{Hada2018}. We note that possible co-spatial jet acceleration and collimation has also been reported for a few other radio galaxies (\citealp[NGC\,6251;][]{Sudou2000}, \citealp[Cygnus\,A;][]{Boccardi2016} and \citealp[NGC\,4261;][]{Yan2023}). Even though, 1928+738 is the first case where an extended ACZ has been discovered in the jet of a FSRQ.

The concurrent acceleration and collimation can be interpreted as a signature of ``differential collimation'' \citep[a.k.a magnetic nozzle effect; e.g.,][]{Camenzind1987, Li1992, Begelman1994}. The mechanism can cause the poloidal magnetic field lines to have both an inner concentration and an outer divergence. This field configuration can act as a nozzle, effectively converting the Poynting flux into kinetic energy \citep[see][]{Komissarov2011, Nakamura2013}. The efficient jet acceleration through the energy conversion gives rise to \emph{linear acceleration}: the bulk Lorentz factor $\Gamma$ increases linearly with jet radius \citep[$\Gamma\,\propto\,R$; e.g.,][]{Beskin2006, Tchekhovskoy2008, Lyubarsky2009}. Again, this matches our observations of 1928+738: the collimation and acceleration profiles (see Section\,\ref{subsec:jetwidth} and \ref{subsec:velfield}) are consistent with $\Gamma \propto R\,(\propto z^{0.46}$) out to $\langle z_{b} \rangle \approx 5.6 \times 10^6\,R_{g}$.

In order to make jet acceleration through differential bunching of poloidal magnetic fields possible, the field lines must be able to communicate laterally \citep[e.g.,][]{Zakamska2008, Tchekhovskoy2009}. This lateral causal connection is maintained if the opening angle of the jet is narrower than that of the Mach cone, meaning $\Gamma\theta_{j} \lesssim {\sigma^{1/2}_{m}}$ \citep{Komissarov2009}, where $\sigma_{m}$ is the magnetization parameter (i.e., the ratio of the Poynting flux and matter energy flux) of the jet. The degree of jet magnetization is still an open problem though, we could infer the lower limit from the simultaneous jet collimation and acceleration. Given that the jet needs to convert the electromagnetic energy into kinetic energy, the jet also needs to be magnetized in the ACZ. Thus we assume $\sigma_{m}\,\gtrsim$ 1; at least moderately magnetized jet. We found that the value of $\Gamma\theta_{j}$ is approximately 0.1\,--\,0.2 along the ACZ of the 1928+738 jet. This implies that the jet satisfies the Mach cone condition. We therefore identify the magnetic nozzle effect as the most likely mechanism for the acceleration and collimation of the 1928+738 jet. Interestingly, the value of $\Gamma\theta_{j}=$\,0.1\,--\,0.2 derived from the ACZ of 1928+738 jet is in good agreement with values for a variety of AGN jets, mostly in distant blazars \citep{Clausen-Brown2013, Jorstad2017,Pushkarev2017}. The jet ACZs in those sources are expected to have been challenging to resolve with previous VLBI observations. The consistency in the $\Gamma\theta_{j}$ values between 1928+738 and the distant blazars, despite the different spatial scales probed, may suggest that the jets possess causal connections over a broad distance range. Furthermore, the termination of the ACZ may not be strongly related to these causal connections but to other factors, such as the environment surrounding the jets, as discussed in Section\,\ref{subsection:soi}.

\subsection{Comparison with Other AGN Sources \label{subsubsection:AGNcomparison}}

\begin{figure*}[t!]
\centering
\includegraphics[width=\textwidth]{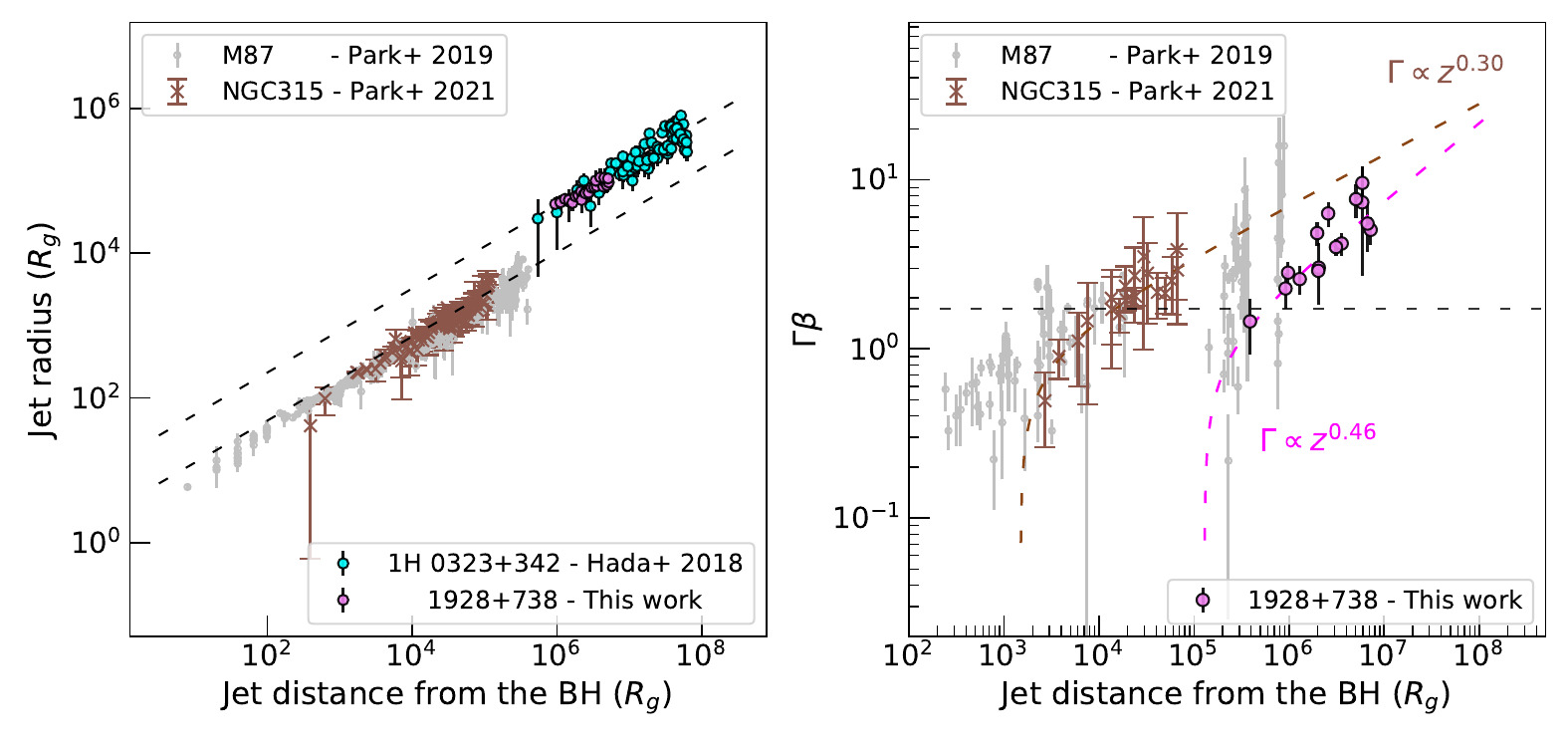}
    \caption{\emph{Left:} Comparison of the jet radius of 1928+738 in the collimation zone as function of deprojected distance from the black hole with M\,87, NGC\,315 and 1H\,0323+342, in units of $R_{g}$. For 1H\,0323+342, $M_{\bullet}=2\times10^{7}\,M_{\odot}$ and $\theta_{v} = 8^{\circ}$ are assumed in this plot. The M\,87 data have been compiled from numerous previous studies \citep[see][and references therein]{Nakamura2018}. The data for NGC\,315 and 1H\,0323+342 are taken from \citet{Park2021} and \citet{Hada2018}, respectively. Their best-fit power-law scaling relations $R \propto z^{0.58}$ are indicated by dashed lines. \emph{Right:} Comparison of the jet velocity field ($\Gamma\beta$) in the acceleration zone as function of deprojected distance from the black hole in units of $R_{g}$ of 1928+738 with those of M\,87 and NGC\,315. The M\,87 data have been compiled from numerous previous studies \citep[see][and references therein]{Park2019-Kine}, and the NGC\,315 data are from \citet{Park2021}. The black dashed horizontal line indicates $\Gamma =2$ for illustration. The best-fit jet acceleration profiles of NGC\,315 \citep{Park2021} and 1928+738 (Section\,\ref{subsec:velfield}) are also presented with colored dashed lines for comparison.
    }
    \label{Fig_4sources}
\end{figure*}

In the following, we compare the jet collimation and acceleration profiles of 1928+738 with those of M\,87, NGC\,315 and 1H\,0323+342. Albeit representing  different AGN types, the jet profiles of other sources in the $10^{3}$\,--\,$10^{5}\,R_{g}$ may provide important information for interpreting the results we find for 1928+738. We note we only regard the jet collimation profile of 1H\,0323+342 because its jet acceleration profile is relatively unclear. The overall results are illustrated in Figure\,\ref{Fig_4sources}.

\subsubsection{Jet Collimation}

In Figure\,\ref{Fig_4sources}, the radii of the jets in collimation zone as function of distance are compared in the left panel. Whereas all jets show similar parabolic geometries, the jet radius profile of 1928+738 is clearly offset from those of M\,87 \citep{Asada2012, Doeleman2012, Hada2013, Nakamura2013, Akiyama2015, Nakamura2018} and NGC\,315 \citep{Park2021}. The profiles of of 1928+738 and 1H\,0323+342 \citep{Hada2018}, however, agree within errors. There is a clear dichotomy: the jets of 1928+738 and 1H\,0323+342 are systematically broader than those of M\,87 and NGC\,315 at any given distance. The best-fit power-law profiles of NGC\,315 and 1H\,0323+342 both show $R\,\propto\,z^{0.58}$ \citep[see][respectively]{Park2021, Hada2018}. Adopting the same power-law index for 1928+738 for comparison (instead of the observed one from Section\,\ref{subsec:jetwidth}), we find a difference in the jet width by a factor $\approx 4$ ($\approx0.6$\,dex).

One explanation for the broad jet of 1928+738 could be a relatively high internal pressure $p_{\rm jet}$, near the black hole. The high initial $p_{\rm jet}$ then results in a relatively large volume (i.e., a broad jet) governed by the pressure equilibrium between the jet and its surroundings \citep{Zakamska2008}. We remark on a similar scenario examined by \citet{Narayan2022}. Their recent simulation at scales of $z\,\leqq\,100\,R_{g}$, suggested that a larger magnetic flux near the central black hole produces a broader jet. If these hold, the magnetic pressure and/or magnetic flux near the black hole of 1928+738 might explain its jet width. Nonetheless, we could not find observational evidence for such a high magnetic flux in 1928+738. \citet{Zamaninasab2014} inferred the magnetic flux threading parsec-scale jets, which is the same as the magnetic flux threading the black hole by the flux-freezing approximation. The inferred magnetic flux of 1928+738 appears comparable to those of M\,87 and other radio galaxies. If this is true, this may indicate that the jets themselves are somewhat similar to each other: then the difference in jet width may be a consequence of the external environment. Indeed, radio-loud quasars (and NLS1s) are widely believed to have a distinctive environment from radio galaxies, such as radiation-driven winds and strong radiation pressure \citep[e.g.,][]{Ohsuga2011, Dorodnitsyn2016, Davis2020}. We note that broad jets have also been found in other FSRQs such as 1633+382 \citep[a.k.a 4C\,38.41;][]{Algaba2019} or 3C\,273 \citep{Okino2022}, although further studies are needed for a quantitative comparison.

\subsubsection{Jet Acceleration}

We present the jet velocity fields $\Gamma\beta$ of 1928+738, M\,87 \citep{Biretta1995, Biretta1999, Cheung2007, Ly2007, Giroletti2012, Meyer2013, Asada2014, Hada2016, Hada2017, Mertens2016, Kim2018, Walker2018, Park2019-Kine} and NGC\,315 \citep{Park2021} in the right panel of Figure\,\ref{Fig_4sources}. Although it is difficult to compare the complex $\Gamma\beta$ profiles in depth, the 1928+738 jet again seems to deviate from the other sources: the velocity profile appears to be shifted outward and/or downward shift, resulting in a relatively low speed at a given distance.

An inward extrapolation of the acceleration profile $\Gamma\,\propto\,z^{0.46}$ suggests that the jet speed remains $v\,\ll\,c$ at distances $z \lesssim 10^{5}\,R_{g}$. The bulk jet velocity of 1928+738 becomes mildly relativistic (e.g., $\Gamma \gtrsim 2$) at distances of a few $10^{5}\,R_{g}$, much later than is the case for the other AGNs or expected theoretically \citep[e.g.,][]{McKinney2006, Penna2013, Nakamura2018}. Given that linear acceleration is only valid when the jet is sufficiently fast \citep[e.g.,][]{Beskin2006, Komissarov2007, Komissarov2009}, one may expect the early-phase acceleration not to be very efficient. If this is the case, the early-phase acceleration profile of the 1928+738 jet in the non-relativistic or sub-relativistic regime could be similar to what has been observed in M\,87\footnote{\citet{Mertens2016} actually suggested a more complicated evolution, including a transition from $\Gamma\,\propto\,z^{0.56}$ to $\Gamma\,\propto\,z^{0.16}$. However, the authors used only the upper envelope of the proper motion data point cloud for deriving the profiles \citep[see Section\,6.2 of][for related discussions]{Park2019-Kine}.} \citep[$\Gamma\,\propto\,z^{0.16}$,][]{Park2019-Kine} or NGC\,315\footnote{We note, however, that \citet{Ricci2022} suggested an efficient acceleration, describing $\Gamma$ as a hyperbolic tangent function of $z$.} \citep[$\Gamma\,\propto\,z^{0.30}$,][]{Park2021}. The jet kinematics at distances of $10^{4}$\,--\,$10^{5}\,R_{g}$ then might give away how the bulk speed of 1928+738 jet remains slow at such far a distance. Moreover, velocity measurements using more densely sampled observations at higher angular resolutions are known to report faster motions in case of the M\,87 jet \citep[see e.g.,][for related discussions]{Walker2008}. Likewise, it may be necessary to perform a jet kinematic analysis using data with higher resolution and observing cadence to obtain a more accurate velocity field of 1928+738. We leave this exploration to future work.

\section{Summary}
\label{sec:summary}

We have investigated the collimation and acceleration of the jet of a nearby FSRQ, 1928+738. We compiled multifrequency and multi-epoch VLBA data from various archives, and reported the observational results. We also performed complementary monitoring observations with KaVA to study the variability of the jet. We summarize our main results as follows:

\begin{enumerate}

\item We obtained multifrequency VLBI images of the 1928+738 jet at various spatial scales, showing an extended jet down to $\approx\,40$\,mas from the radio core. Based on a 2D cross-correlation analysis of optically thin jet structure, we measured the frequency-dependent location of the radio core, which follows $z_{c} \propto \nu^{-1/k_{z}}$, where $k_{z}=1.87\pm0.48$. The core-shift measurements allow us to accurately align the jet collimation and acceleration profiles obtained from the multifrequency data.

\item We extensively investigated the jet radius profile, covering the distance range from $\sim0.8$\,mas to $\sim40$\,mas. We discovered that the jet geometry transits from a parabolic shape ($R\,\propto\,z^{0.46\pm0.10}$) to a conical shape ($R\,\propto\,z^{1.02\pm0.03}$), at a distance of $z_{b}^{w}\,=\,4.72\pm0.72$\,mas. We also performed a complementary kinematic analysis. We found that the bulk Lorentz factor $\Gamma$, too, transits from an acceleration ($\Gamma \propto z^{0.46\pm0.08}$) to a deceleration ($\Gamma \propto z^{-0.26\pm0.25}$) at a distance of $z_{b}^{\Gamma}\,=\,5.39\pm1.41$\,mas. By combining the collimation profile and the jet velocity field, we discovered that jet collimation and acceleration occur in the same region.

\item We studied the jet variability using our monitoring observations with KaVA at 43\,GHz. We derived the variability Doppler factors $\delta_{\rm var}$ and the viewing angle $\theta_{\rm v}$, for four knots at distances $\lesssim5$\,mas, which are reliably cross-identified throughout the $\approx2$\,years of observation. We found that the viewing angle remains constant along the jet within errors, with an average value $\langle \theta \rangle = 13.2^\circ$.

\item Adopting a viewing angle $\theta_{\rm v} = 13.2^{\circ}$ and a black hole mass $M_{\bullet} \approx 3 \times 10^{8}\,M_{\odot}$, the deprojected distance to the downstream end of the ACZ is $\langle z_{b} \rangle \approx 5.6 \times 10^6\,R_{g}$. This is similar to the sphere of gravitational influence of the black hole in the center of 1928+738. This implies that the physical extent of the ACZ in the 1928+738 jet is governed by gravitational field of the black hole. The jet might be collimated by winds from accretion flows.

\item We found that the jet gradually accelerates to highly relativistic speeds with $\Gamma\,\sim\,10$ over a distance of several million $R_{g}$. We found a linear relation between bulk Lorentz factor and jet radius, $\Gamma\,\propto\,R$, in agreement with theoretical expectations for MHD acceleration and collimation of highly magnetized jets. We conclude that the jet is accelerated through the magnetic nozzle effect, efficiently converting the Poynting flux into kinetic energy.

\item We compared the jet collimation and acceleration of 1928+738 with other AGNs. The jet of 1928+738 is broader at a given distance than those of M\,87 and NGC\,315. We enumerate two possibilities: (i) the magnetic pressure and/or magnetic flux near the black hole of 1928+738 might be substantially higher than in the other sources; (ii) the broad jet might be a consequence of the distinctive external environment of the 1928+738 jet compared to the jets from radio galaxies. In addition, we found that the bulk jet speed of 1928+738 is relatively slow at a given distance. This implies that the early-phase acceleration of the jet is much flatter than $\Gamma\,\propto\,z^{0.46}$ at distances $\leq\,10^{5}\,R_{g}$. 

\end{enumerate}

\begin{acknowledgements}

We thank the anonymous referee for constructive and detailed comments, which were helpful to improve the manuscript. We acknowledge the use of data from the Astrogeo Center database maintained by Leonid Petrov, Yuri Kovalev and Yuzhu Cui. This research has made use of data from the MOJAVE database that is maintained by the MOJAVE team \citep{Lister2018}. The VLBA is an instrument of the National Radio Astronomy Observatory. The National Radio Astronomy Observatory is a facility of the National Science Foundation operated by Associated Universities, Inc. This work is made use of the East Asian VLBI Network (EAVN), which is operated under cooperative agreement by National Astronomical Observatory of Japan (NAOJ), Korea Astronomy and Space Science Institute (KASI), with the operational support by Kagoshima University (for the operation of VERA Iriki antenna). We acknowledge financial support from the National Research Foundation of Korea (NRF) through grant 2022R1F1A1075115. This work was supported by the Ministry of Education of the Republic of Korea and the National Research Foundation of Korea (BK21 FOUR).

\end{acknowledgements}

\bibliographystyle{aa}
\bibliography{main.bib}

\begin{appendix} 
\label{appendix}

\section{\emph{Modelfit} Jet Kinematics \label{sec:appendix_kin}}

\begin{figure*}[t]
\centering
\includegraphics[width=\textwidth]{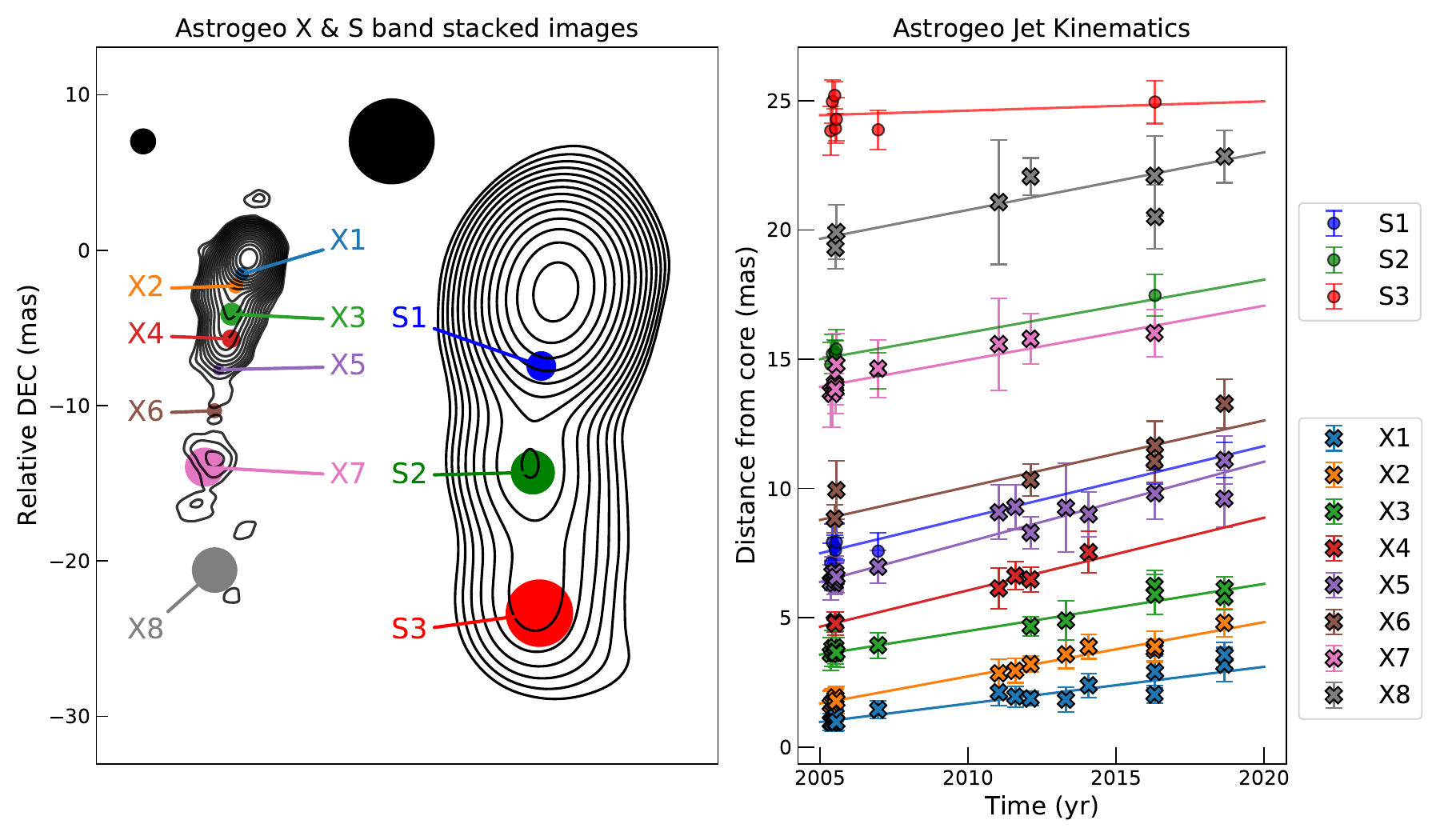}
    \caption{\emph{Left:} Stacked images of naturally weighted CLEAN maps of 1928+738 at X- and S-band from the Astrogeo program. The averaged circular beams with sizes of 1.57~mas and 5.43~mas, respectively, are shown to the top left of each contour map. The jet components are labeled individually. \emph{Right:} Distance from the core as function of time (data points) for all Astrogeo knots detected at $\lesssim$ 8.6\,GHz, with linear fits (straight lines) overlaid. Different markers ares used to distinguish observing frequencies.}
    \label{Fig:Astrogeo_kin}
\end{figure*}

\begin{figure*}[t]
\centering
\includegraphics[width=\textwidth]{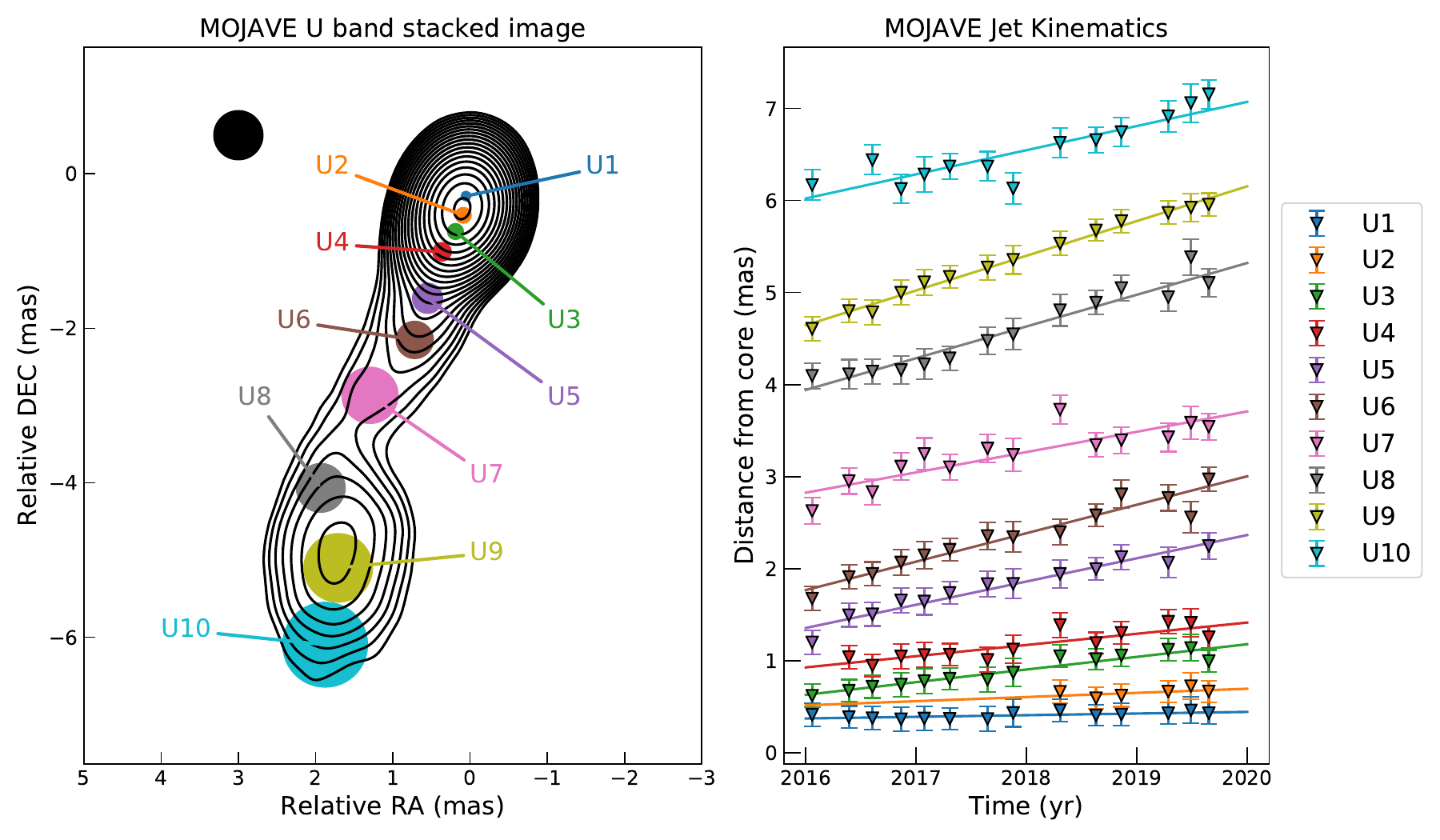}
    \caption{\emph{Left:} Stacked image of naturally weighted CLEAN maps of 1928+738 at U band from the MOJAVE program. The averaged circular beam with a size of 0.63~mas is shown in the top left corner. Jet components are labeled individually. \emph{Right:} Distance from the core as function of time (data points) for the MOJAVE knots detected at 15\,GHz, with linear fits (straight lines) overlaid.}
    \label{Fig:MOJAVE_kin}
\end{figure*}

We provide here additional information about our kinematic analysis. The Astrogeo X- and S-band data sets required more careful analysis compared to the MOJAVE and KaVA data sets, due to their non-uniform angular resolution and cadence. Thus we only analyzed the knots located at least about one (averaged) beam away from the core, in order to avoid mis-identifications. We cross-identified in total 8 knots at X- and 3 at S-band. The knots detected at X-band are labeled X1, X2, ..., and X8 while the knots detected at S-band are labeled S1, S2, and S3. We also cross-identified 10 knots over time from the MOJAVE U\,band data set and labeled them U1, U2, U3, ..., and U10. We present the knot locations within stacked Astrogeo and MOJAVE images in Figure\,\ref{Fig:Astrogeo_kin} and Figure~\ref{Fig:MOJAVE_kin}, respectively. We note that the CLEAN images are stacked after convolution with the average circular beams. The motion of the knots with time is presented along the maps. In Table~\ref{table:Astrogeo_params} and Table~\ref{table:Mojave_params}, the mean radial distance $\langle z \rangle$, angular proper motion (in mas/yr), and the apparent speed (in units of light speed) of individual knots are listed.

\begin{table}[h]
\centering
\caption{Astrogeo Jet Kinematics
 \label{table:Astrogeo_params}}
\begin{tabular}{cccc}
\hline
\hline
Knot ID & $\langle \rm z \rangle$ (mas) & $\rm \mu_{z}$ (mas $\rm yr^{-1})$ & $\rm \beta_{app}$ (c) \\

\hline
X1 & 1.84  & 0.14 $\pm$\ 0.02 & 2.70 $\pm$\ 0.38 \\
X2 & 2.82  & 0.21 $\pm$\ 0.03 & 3.98 $\pm$\ 0.48  \\
X3 & 4.60  & 0.18 $\pm$\ 0.03 & 3.47 $\pm$\ 0.48 \\
X4 & 6.07  & 0.28 $\pm$\ 0.06 & 5.34 $\pm$\ 1.20  \\
X5 & 8.11  & 0.31 $\pm$\ 0.04 & 5.89 $\pm$\ 0.81 \\
X6 & 10.85 & 0.26 $\pm$\ 0.06 & 4.87 $\pm$\ 1.16  \\
X7 & 14.60 & 0.21 $\pm$\ 0.08 & 3.98 $\pm$\ 1.54  \\
X8 & 21.11 & 0.22 $\pm$\ 0.08 & 4.24 $\pm$\ 1.50  \\ \hline
S1 & 8.50  & 0.28 $\pm$\ 0.05 & 5.25 $\pm$\ 0.90  \\
S2 & 15.41 & 0.20 $\pm$\ 0.08 & 3.90 $\pm$\ 1.49  \\
S3$^{*}$ & 24.43 & 0.04 $\pm$\ 0.08 & 0.68 $\pm$\ 1.56  \\ \hline
\multicolumn{4}{l}{$^{*}$ Not included into analysis, due to large uncertainty.}\\
\end{tabular}

\end{table}

\begin{table}
\centering
\caption{MOJAVE Jet Kinematics
 \label{table:Mojave_params}}
\begin{tabular}{cccc}
\hline
\hline
Knot ID & $\langle \rm z \rangle$ (mas) & $\rm \mu_{z}$ (mas $\rm yr^{-1})$ & $\rm \beta_{app}$ (c) \\

\hline
U1$^{*}$ & 0.40  & 0.02 $\pm$\ 0.03 & 0.34 $\pm$\ 0.55  \\
U2$^{\dag}$ & 0.65  & 0.05 $\pm$\ 0.10 & 0.86 $\pm$\ 1.99  \\
U3 & 0.88  & 0.14 $\pm$\ 0.03 & 2.61 $\pm$\ 0.55  \\ 
U4 & 1.17  & 0.12 $\pm$\ 0.03 & 2.31 $\pm$\ 0.63 \\
U5 & 1.79  & 0.25 $\pm$\ 0.03 & 4.80 $\pm$\ 0.66  \\
U6 & 2.33  & 0.31 $\pm$\ 0.03 & 5.87 $\pm$\ 0.60 \\
U7 & 3.24  & 0.22 $\pm$\ 0.03 & 4.19 $\pm$\ 0.65 \\
U8 & 4.58  & 0.34 $\pm$\ 0.04 & 6.55 $\pm$\ 0.67 \\
U9 & 5.34  & 0.38 $\pm$\ 0.03 & 7.16 $\pm$\ 0.57  \\
U10 & 6.54 & 0.26 $\pm$\ 0.04 & 4.98 $\pm$\ 0.78  \\ \hline
\multicolumn{4}{l}{$^{*}$ Not included into analysis, due to large uncertainty.}\\
\multicolumn{4}{l}{$^{\dag}$ Not included into analysis, due to large uncertainty.}\\
\end{tabular}

\end{table}

\FloatBarrier

\section{Outlier Identification \label{sec:kin_outlier}}

\begin{figure*}[b]
\centering
\includegraphics[width=\textwidth]{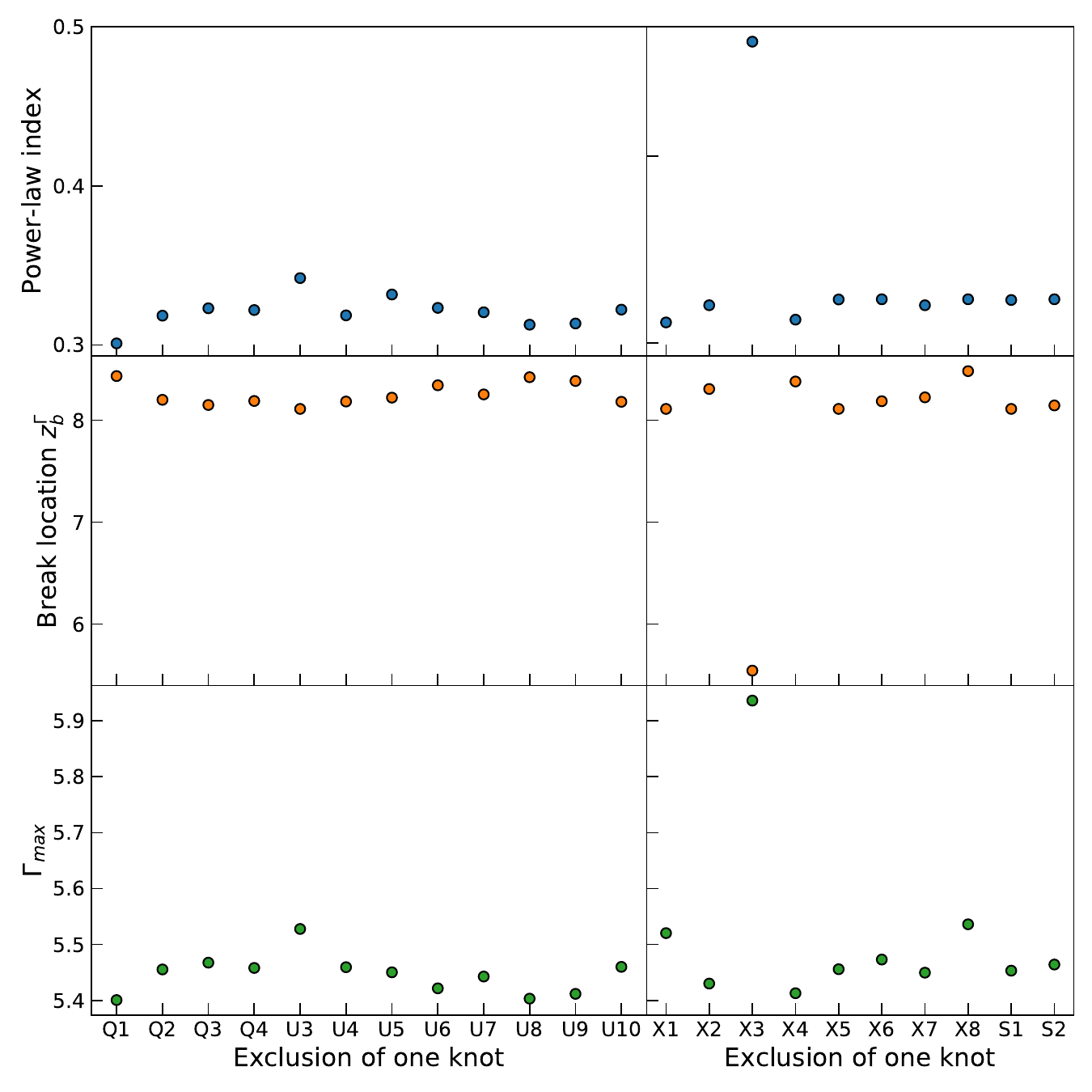}
    \caption{Best-fit parameter values, from top to bottom: Power-law index, break location $z_{b}^{\Gamma}$, and maximum Lorentz factor $\Gamma_{\rm max}$, for each bootstrap sample. Each bootstrap sample includes all data except for the knot named at the abscissa.}
    \label{Fig:outlier}
\end{figure*}

As described in Section\,\ref{subsec:velfield}, we model the bulk Lorentz factor of 1928+738 as function of distance with a broken power-law function. However, one of the knots is identified as an outlier and excluded from the fit as follows. We generate multiple bootstrap samples, fit a broken power-law model to each sample, and then examine the stability of the best-fit model parameters: the power-law index, the break location $z_{b}^{\Gamma}$, and the maximum Lorentz factor $\Gamma_{\rm max}$. Since the number of knots is small, each bootstrap sample includes all data points except one. In Figure\,\ref{Fig:outlier}, we the best-fit parameter values for each bootstrap sample. Knot X3 is an obvious outlier.

\FloatBarrier


\section{Slow jet speeds and the Counter-Jet} \label{sec:cj_speed}

\begin{figure}[t!]
\centering
\includegraphics[width=0.47\textwidth]{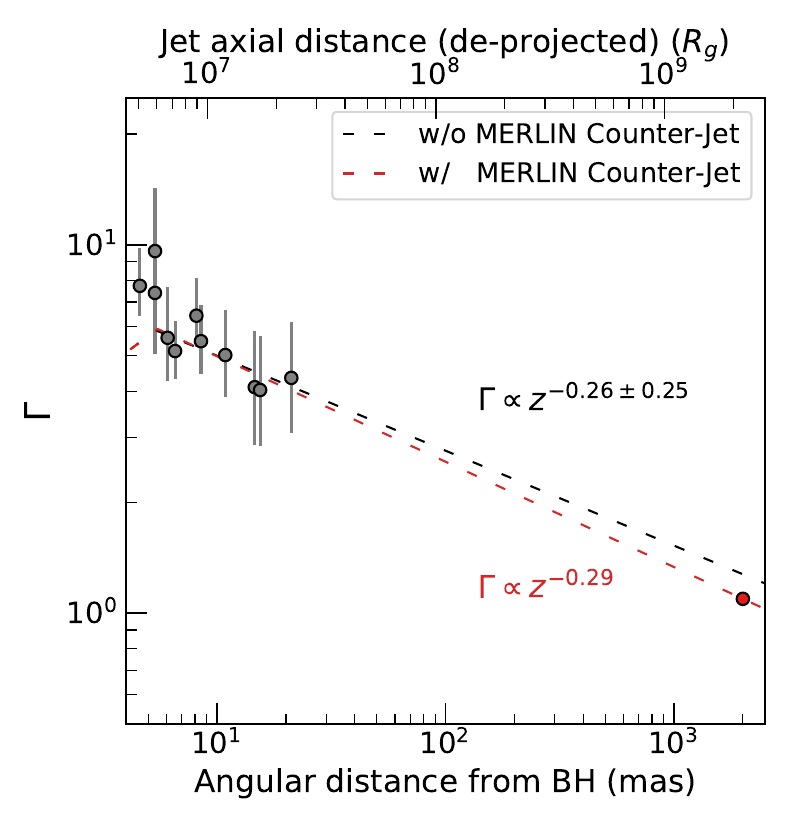}
    \caption{The bulk Lorentz factor as a function of distance $z$ from the black hole. The distance is displayed in units of $R_{g}$ (top axis) and mas (bottom axis). The data includes the decelerating part in Figure\,\ref{Fig:AZ} (gray data points), and the result inferred from the MERLIN observation at 408\,MHz (red data point; see text). Error bars represent $\pm\,1\,\sigma$ uncertainties. The black dashed line represents the best-fit broken power-law model in Figure\,\ref{Fig:AZ}, and the red one represents the model obtained by including the red data.
}
    \label{Fig:DZ}
\end{figure}

In this appendix, we discuss the counter-jet of 1928+738 and its detectability. As introduced in Section\,\ref{sec:intro}, it is evident that 1928+738 shows two-sided morphology on arcsecond scales \citep[e.g.,][]{Johnston1987, Hummel1992}. The difference in the brightness of the two-sided jets at the same distance can be explained by the result of Doppler boosting and de-boosting, assuming that they are intrinsically the same. Then the jet-to-counter-jet brightness ratio ($BR$) of is directly related to the bulk flow speed $\beta$ and the viewing angle $\theta_{\rm v}$, via

\begin{equation}\label{cj}
\centering
        BR \equiv \frac{I_{\rm jet}}{I_{\rm cjet}} = \left(\frac{1+\beta\,{\rm cos\,\theta_{v}}}{1-\beta\,{\rm cos\,\theta_{v}}}\right)^{2-\alpha},
\end{equation}

\noindent where $I_{\rm jet}$ and $I_{\rm cjet}$ denote the intensities of the jet and counter-jet at the same distance, and $\alpha$ is the spectral index. Adopting the $BR\,\approx\,10$, which is measured by a MERLIN 408\,MHz observation at 2\,arcsec \citep{Hummel1992}, we could infer the jet speed at such a far distance. We present the result in Figure\,\ref{Fig:DZ} as a form of the bulk Lorentz factor $\Gamma$, together with a part of the velocity field in Figure\,\ref{Fig:AZ}, where shows a gradual deceleration. The jet speed inferred from the $BR$ shows a good agreement with other data points obtained from the kinematics analysis. We found that the broken power-law fit of the velocity field (Section\,\ref{subsec:velfield}) is not greatly affected by an inclusion of this new data point (see Figure\,\ref{Fig:DZ}). This may imply that there is indeed a jet deceleration and it is maintained down to over $10^{9}\,R_{g}$ however, it is not used for our main analysis owing to the large distance gap (i.e., $\approx$\,2 orders of magnitude).

On the other hand, there is little evidence of the counter-jet in VLBI observations on (sub)-mas scales (see Figure\,\ref{Fig:imgs}). Nevertheless, the non-detection also allows us to constrain the jet speed at the nearest distance of the proper motion. As described in Section\,\ref{subsec:KaVA_var}, the slowest motion is detected approximately at 0.35\,mas from 43\,GHz KaVA observations. At the location, we can estimate $BR\,\gtrsim\,435$ from the 43\,GHz KaVA image (see Figure\,\ref{Fig:imgs}), by adopting $I_{\rm cjet}\,\lesssim\,5\,\sigma_{\rm rms}$. We could infer the lower limit to the jet speed $\Gamma\,\gtrsim\,1.67$ for the case $\alpha\,=\,-0.5$, which agrees well with the slowest speed detected (i.e., $\Gamma$\,=\,1.76). This supports the robustness of our main analysis. We also note that a slightly different value of $\alpha$ between -1 and 0 does not affect our conclusion. We plan to have a more robust analysis on the counter-jet with future multifrequency observations with higher angular resolution and higher sensitivity (K. Yi et al. 2024, in preparation).

\end{appendix}

\end{document}